\documentclass[journal]{IEEEtran}
\usepackage{amsmath}
\usepackage{cite}
\usepackage{amssymb}
\usepackage{array}
\usepackage{amsfonts}
\usepackage[ruled]{algorithm2e}
\usepackage{autobreak}
\usepackage{multirow}
\usepackage{booktabs}
\usepackage[cmyk]{xcolor}
\usepackage{comment}
\usepackage{dsfont}
\usepackage{graphicx}
\usepackage{epsfig}
\usepackage{subfigure}
\usepackage{float}
\usepackage{epstopdf}
\usepackage{psfrag}
\usepackage{xcolor}
\usepackage{url}
\usepackage{tabularx}
\usepackage[colorlinks,linkcolor=black,urlcolor=black,anchorcolor=black,citecolor=black,hyperfootnotes=true]{hyperref}

\newtheorem{remark}{Remark}

\usepackage{mathtools}

\title{$\text{m}^3$TrackFormer: Transformer-based mmWave Multi-Target Tracking with Lost Target Re-Acquisition Capability}

\author{\IEEEauthorblockN{Tongkai Li, Weifeng Zhu,
Shuowen Zhang,
Jiannong Cao, Shuguang Cui, 
and Liang Liu}

\thanks{An earlier version of this paper was presented in part at the 2026 IEEE International Conference on Acoustics, Speech, and Signal Processing (ICASSP) \cite{tongkai_2026_ICASSP}.}

\thanks{T. Li, W. Zhu, S. Zhang, and L. Liu are with the Department of Electrical and Electronic Engineering, The Hong Kong Polytechnic University, Hong Kong SAR, China (e-mails: tongkai.li@connect.polyu.hk, \{eee-wf.zhu, shuowen.zhang, liang-eie.liu\}@polyu.edu.hk).}
\thanks{J. Cao is with the Department of Computing, The Hong Kong Polytechnic University, Hong Kong SAR, China (e-mail: jiannong.cao@polyu.edu.hk).}
\thanks{S. Cui is with the School of Science and Engineering and the Future
Network of Intelligence Institute, The Chinese University of Hong Kong, Shenzhen, 518172, China (e-mail: shuguangcui@cuhk.edu.cn).}
}

\begin{document}
\maketitle

\begin{abstract}
This paper considers a millimeter wave (mmWave) integrated sensing and communication (ISAC) system, where a base station (BS) equipped with a large number of antennas but a small number of radio-frequency (RF) chains emits pencil-like narrow beams for persistent tracking of multiple moving targets. Under this model, the tracking lost issue arising from the misalignment between the pencil-like beams and the true target positions is inevitable, especially when the trajectories of the targets are complex, and the conventional Kalman filter-based scheme does not work well. To deal with this issue, we propose a Transformer-based mmWave multi-target tracking framework, namely $\text{m}^3$TrackFormer, with a novel re-acquisition mechanism, such that even if the echo signals from some targets are too weak to extract sensing information, we are able to re-acquire their locations quickly with small beam sweeping overhead. Specifically, the proposed framework can operate in two modes of normal tracking and target re-acquisition during the tracking procedure, depending on whether the tracking lost occurs. When all targets are hit by the swept beams, the framework works in the Normal Tracking Mode (N-Mode) with a Transformer encoder-based Normal Tracking Network (N-Net) to accurately estimate the positions of these targets and predict the swept beams in the next time block. While the tracking lost happens, the framework will switch to the Re-Acquisition Mode (R-Mode) with a Transformer decoder-based Re-Acquisition Network (R-Net) to adjust the beam sweeping strategy for getting back the lost targets and maintaining the tracking of the remaining targets. Thanks to the ability of global trajectory feature extraction, the $\text{m}^3$TrackFormer can achieve high beam prediction accuracy and quickly re-acquire the lost targets, compared with other tracking methods. Numerical experiments demonstrate that the $\text{m}^3$TrackFormer can maintain high tracking success probability with much longer tracking durations than the representative benchmarks.

\end{abstract}
\begin{IEEEkeywords}
Integrated sensing and communication (ISAC), multi-target tracking, millimeter wave, Transformer, target re-acquisition.
\end{IEEEkeywords}

\vspace{-0.2cm}
\section{Introduction}
\subsection{Motivation}
\begin{figure}[t]
    \centering
    \includegraphics[width=0.95\linewidth]{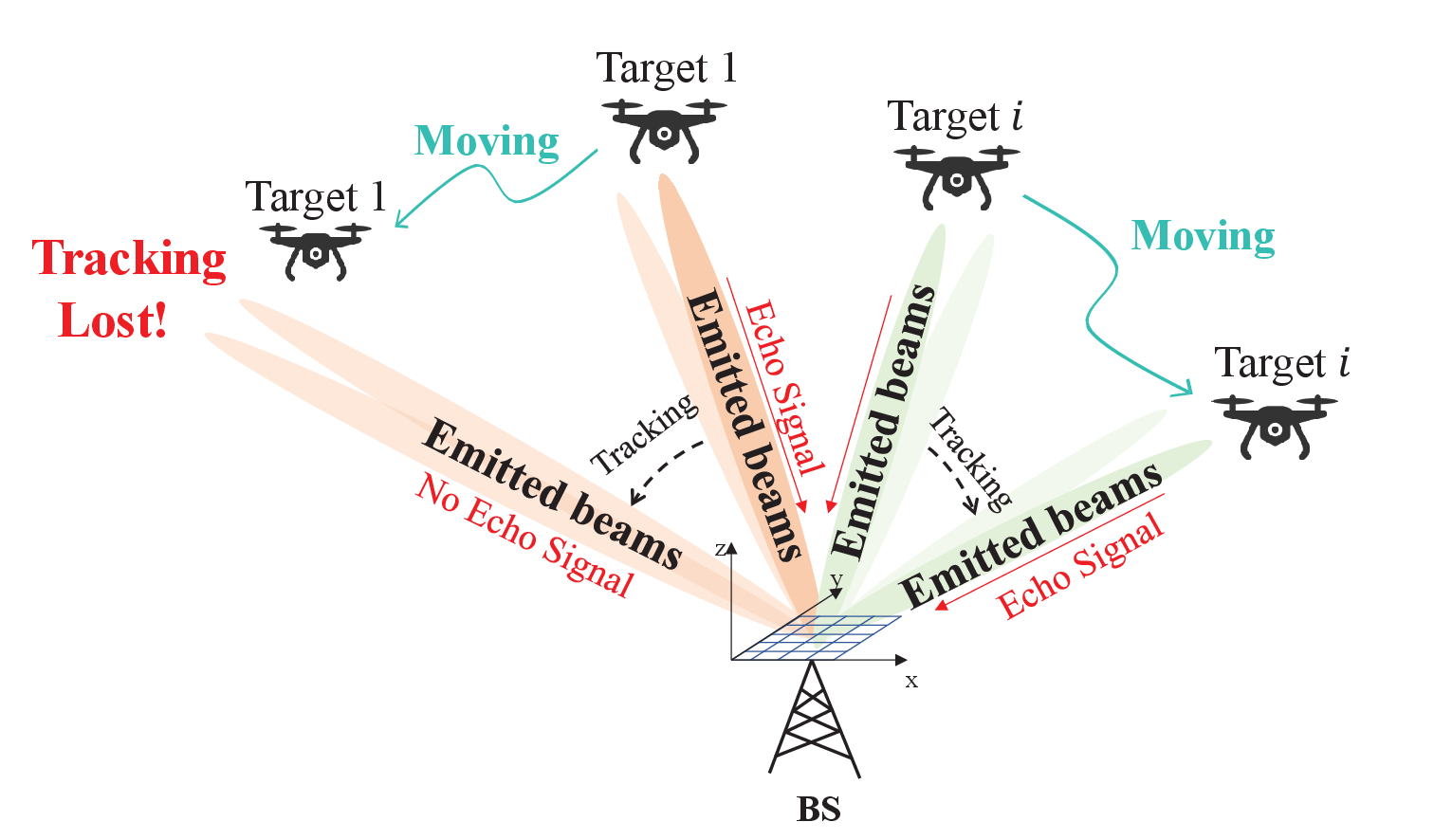}
\vspace{-0.2cm}
    \caption{System model for target tracking in the 6G mmWave ISAC system: tracking lost event occurs when the pencil-like mmWave beams are not precisely pointed to the target positions.}
    \label{fig:model}
    \vspace{-0.5cm}
\end{figure}

ITU-R has recently identified integrated sensing and communication (ISAC) as a primary usage scenario of the sixth-generation (6G) cellular network \cite{IMT}. This inspires a great amount of effort in investigating the integration of sensing functionality into a communication-oriented cellular network \cite{isac1,isac2,isac3,isac4}. Notably, the millimeter wave (mmWave) band promised in 6G network provides sufficient bandwidth that is beneficial to most of the sensing applications, e.g., high-resolution localization and imaging. However, the story is totally different to tracking, which aims at precisely localizing moving targets at each time block based on historical signals and is essential for many applications such as beam alignment in mobile mmWave communication \cite{beamalignment} and unregistered drone detection in low-altitude economy \cite{lae}. Specifically, due to the expensive radio frequency (RF) chains at the mmWave band, analog beamforming is widely adopted in mmWave systems, which yields narrow beams. The pencil-like narrow beams make the 6G mmWave base station (BS) lack adequate historical signals for tracking, because some of its previously emitted beams may not hit the moving targets. This gives rise to the tracking lost issue \cite{heath}, as shown in Fig. \ref{fig:model}. 

There are two important actions for tackling the tracking lost issue in 6G mmWave systems: prevention and re-acquisition. \textbf{Action 1 - Prevention}: in a time block, if the beams are well aligned with the target, i.e., the target is not lost currently, we should precisely predict the location of this target for beam alignment in the next time block such that the target is not lost with high probability in the future block. \textbf{Action 2 - Re-acquisition}: in a time block, if the beams are not well aligned with the target such that no sensing information can be extracted from the received signal, we should re-acquire the target location in the future blocks via transmitting beams to various sites. 

In the literature, the classic approaches for prevention and re-acquisition in mmWave tracking systems are Kalman Filter (KF) and exhaustive beam sweeping, respectively. The KF leverages the state space model and the available measurements to predict the future location of the target \cite{ekf}. However, its performance is sensitive to the accuracy of this prior knowledge and state measurements in each time block, which may be severely degraded with inaccurate state model and heavily corrupted measurements \cite{kalmannet}. The exhaustive beam sweeping performs re-acquisition by scanning beams over the whole space to re-identify the location of the target. However, this approach will take so much time in practice, and will generate interference to communication users at all positions. Recently, deep learning (DL)-based solutions to mmWave tracking have emerged as promising alternatives. The DL methods aim to learn the inherent dynamics of the target mobility from the pre-collected data, thereby overcoming the deficiency of explicit characterization of the complicated mobility model. Therefore, they are well suitable to tackle the prevention and re-acquisition tasks in challenging tracking scenarios involving highly nonlinear and complex target trajectories. 

\subsection{Prior Works}
Recently, the DL-based tracking problem in mmWave systems has been investigated in \cite{ kalmannet,pedraza2025HMMTWC,liu2022learningJSAC, Shah2022AutoEncoderLSTM,yuan2020learning, Zhang2020Bandit,Zhang2020Intelligent,Susarla2022Learning}, which mainly focus on the prevention issue. Specifically, in \cite{kalmannet}, a hybrid model-based and data-driven method is proposed to learn the filtering operation of KF by a recurrent neural network (RNN), thereby enhancing the tracking performance in the case with model mismatch. Similarly, \cite{pedraza2025HMMTWC} addresses the codebook-based analog beam tracking problem by introducing a Hidden Markov Model (HMM)-based filter realized by a deep neural network (DNN) to learn the transition probabilities between candidate beam states, which eliminates the need for prior knowledge of the mobility model. In \cite{liu2022learningJSAC}, a long short-term memory (LSTM) network is designed to directly predict the beamforming matrix from the extracted spatial features output by the Convolutional Neural Network (CNN) in the vehicular network, thus bypassing the need for explicit channel tracking. For the multi-cell system, \cite{Shah2022AutoEncoderLSTM} proposes an autoencoder-LSTM architecture to jointly predict multiple beams across different cells with reduced computational complexity. \cite{yuan2020learning} proposes a LSTM-based predictive beamforming algorithm for UAV communications, where real-time angle refinement is incorporated to mitigate beam misalignment caused by UAV jittering. Despite their effectiveness, RNN- and LSTM-based models exhibit limited capability in capturing globally temporal dependencies due to their sequential processing mechanism, thus suffer from low-precision tracking performance for targets with complicated mobilities. Beyond the above supervised learning methods, reinforcement learning (RL) approaches such as multi-armed bandit (MAB) and Deep Q-learning have also been proposed to tackle the beam tracking problem by interacting with the environment in real time to learn an optimal beam prediction policy \cite{Zhang2020Bandit,Zhang2020Intelligent,Susarla2022Learning}. Compared with RNN-based schemes, RL-based approaches exhibit stronger generalization capability in complex and dynamic environments. However, in mmWave tracking scenarios where the number of candidate beams is large, the RL-based approaches often suffer from slow convergence speed and time-consuming training process due to the high dimensionality of the action and state space. 

Despite the various advances on prevention, the tracking lost remains as a critical problem in the mmWave system. In the case with tracking lost, conventional DL-based approaches in \cite{kalmannet,pedraza2025HMMTWC,liu2022learningJSAC, Shah2022AutoEncoderLSTM,yuan2020learning} fail to work due to the missing measurements on the tracking targets \cite{transformer_traj}. In practice, the rapid trajectory variations can easily result in tracking lost. As a solution, the data imputation technique can be employed with existing tracking schemes to estimate these missing entries and predict the swept beams in the future \cite{Che2018RNNMissing}. However, this method ignores the trajectory variations of the lost targets for beam prediction, resulting in severe error accumulation \cite{Du2023SAITS,Venkatraman2015MultiStep} and quite low probability of lost target re-acquisition. Consequently, the frequent exhaustive beam sweeping operation is required for finding the lost targets. To avoid the reliance on exhaustive beam sweeping, this work is motivated to design a robust target tracking scheme enabled by a complementary target re-acquisition mechanism with low beam sweeping overhead. To the best of our knowledge, the research on the tracking problem with the consideration of lost targets re-acquisition is still missing, and this work is the first attempt to address this problem.

\subsection{Main Contributions}

This paper aims to design a robust multi-target tracking framework in the 6G mmWave ISAC system. Motivated by the powerful capability of the Transformer architecture in time-series data processing \cite{transformer_time_series,jiang}, we propose a novel Transformer-based framework to simultaneously track multiple targets and predict the swept beams in the next time block. In particular, the proposed framework can facilitate fast re-acquisition of lost targets by exploiting the temporal dependencies in incomplete trajectory sequences and the negative tracking lost events. The main distinctions and contributions of this work can be summarized as follows: 
\begin{itemize}
    \item This paper proposes a two-mode Transformer-based framework for the mmWave multi-target tracking task. Specifically, the proposed tracking framework executes its Normal Tracking Mode (N-Mode) to perform tracking on all targets if no target lost events occur. Otherwise, the Re-Acquisition Mode (R-Mode) is triggered to get back the lost targets as soon as possible. In contrast to the conventional tracking methods, the proposed framework can realize robust target tracking performance with low beam sweeping overhead by benefiting from the carefully-designed R-Mode.
    \item We first propose a novel $\text{m}^2$TrackFormer in the single-target scenario, which consists of a Normal Tracking Network (N-Net) and a Re-Acquisition Network (R-Net). Specifically, the N-Net is designed based on the Transformer encoder architecture and activated in the N-Mode to perform tracking and beam prediction, which leverages the powerful Self-Attention mechanism to capture the global motion features during the tracking process. Then, the R-Net is designed based on the Transformer decoder architecture and utilized for the R-Mode, which applies the Cross-Attention mechanism to integrate the negative information from beam misalignment and positive information in historical trajectory for target re-acquisition. Compared with traditional tracking schemes, the implicit information in the target-beam misalignment is also leveraged in the tracking procedure, thereby enhancing the tracking robustness.
    \item  We then propose a scalable $\text{m}^3$TrackFormer in the multi-target scenario to support the tracking and re-acquisition of multiple targets. Compared with $\text{m}^2$TrackFormer, the motion features of each target are extracted in parallel and then aggregated to generate a joint beam sweeping strategy in $\text{m}^3$TrackFormer, which realizes flexible adjustment of the swept beams with an arbitrary number of lost targets. 
    \item  Numerical results demonstrate that the proposed framework consistently achieves high tracking success probability with much longer tracking durations than conventional KF and RNN-based tracking schemes. Specifically, the tracking duration is improved by 15\% in low-mobility scenarios and by more than 130\% in high-mobility scenarios, which validates the effectiveness of the proposed re-acquisition mechanism in handling rapid trajectory variations and re-acquiring lost targets without introducing extra beam resources. Moreover, the framework maintains millisecond-level inference latency as the number of targets increases, demonstrating its scalability and feasibility for real-time implementation in 6G mmWave ISAC systems. 
\end{itemize}
\vspace{-0.2cm}
\subsection{Organization}
The rest of the paper is organized as follows. Section II introduces the system model of the 6G mmWave tracking system. Section III formulates the problem and introduces the two-mode tracking framework used to solve it. Section IV provides a detailed illustration of the network design for the proposed framework in the single-target scenario. The general tracking framework for the multi-target scenario is introduced in Section V. Finally, Section VI evaluates the performance of the proposed tracking framework, and Section VII concludes this work. 

\textit{Notation}: Column vectors and matrices are denoted by boldfaced lowercase and uppercase letters, e.g., $\mathbf{x}$ and $\mathbf{X}$. $\mathbb{R}^{n \times n}$ and $\mathbb{C}^{n \times n}$ represent the sets of $n$-dimensional real and complex matrices, respectively. The superscripts $(\cdot)^T$ and $(\cdot)^H$ denote the transpose and conjugate transpose operations, respectively. $U[a,b]$ denotes the probability density function of the uniform distribution on the interval $[a,b]$. $\mathbf{x}[i]$ denotes the $i$-th element of the vector $\mathbf{x}$. $\mathbf{X}[i,:]$, $\mathbf{X}[:,j]$, and $\mathbf{X}[i,j]$ denote the $i$-th row, the $j$-th column, and the $(i,j)$-th element of the matrix $\mathbf{X}$, respectively. $\mathbf{I}_n$ denotes an $n \times n$ identity matrix. $\mathbf{0}_n$ and $\mathbf{1}_n$ denote an all-zero vector and all-one vector of length $n$, respectively. $\lVert \cdot \rVert_2$ denotes the Euclidean norm. $\lvert \cdot \rvert$ denotes the magnitude of a complex scalar.
\vspace{-0.1cm}
\section{System Model}

We consider a mmWave orthogonal frequency division multiplexing (OFDM) ISAC system in which a BS transmits radio-frequency signals for tracking $I$ moving targets, denoted by $\mathcal{I}=\{1,\ldots,I\}$, based on their echo signals, and communicating with $U$ users, denoted by $\mathcal{U}=\{1,\ldots, U\}$. The BS is equipped with a transmit uniform planar array (UPA) of \(N_\text{T}=N_\text{T}^x \times N_\text{T}^y\) antennas and $N_{\text{T}}^{\text{RF}}\ll N_{\text{T}}$ radio frequency (RF) chains, and a receive UPA of \(N_{\text{R}}=N_\text{R}^x \times N_\text{R}^y\) antennas and $N_{\text{R}}^{\text{RF}} \ll N_{\text{R}}$ RF chains. Denote the BS location as \(\mathbf{p}_{\mathrm{BS}}=[x_{\mathrm{BS}}, y_{\mathrm{BS}}, z_{\mathrm{BS}}]^T\) in a three-dimensional (3D) Cartesian coordinate system. The targets move over a time duration of $T$ seconds (s), which consists of \(Q\) time blocks, each with a duration of $\Delta T=T/Q$~s. The location of each target is assumed to be fixed within one block, but varying at different blocks \cite{Zhu2025RIS_ISAC,Tichavsky1998PCRB}. In the $q$-th block, the location of target $i\in\mathcal{I}$ is denoted as \(\mathbf{u}_{i,q}=[x_{i,q}, y_{i,q}, z_{i,q}]^T\), \(q=1,\ldots, Q\), while the range, azimuth angle, and elevation angle of target $i$ relative to the BS are denoted by $d_{i,q}=
\left\| \mathbf{u}_{i,q} - \mathbf{p}_{\mathrm{BS}} \right\|_2$, $\theta_{i,q}=
\mathrm{arctan}\left(\frac{
y_{i,q}-y_{\mathrm{BS}}}{
x_{i,q}-x_{\mathrm{BS}}}
\right)$, and $\phi_{i,q}=
\arccos\!\left(
\frac{z_{i,q}-z_{\mathrm{BS}}}
{d_{i,q}}
\right)$, respectively.  

\begin{figure}[t]
    \centering
    \includegraphics[width=0.98\linewidth]{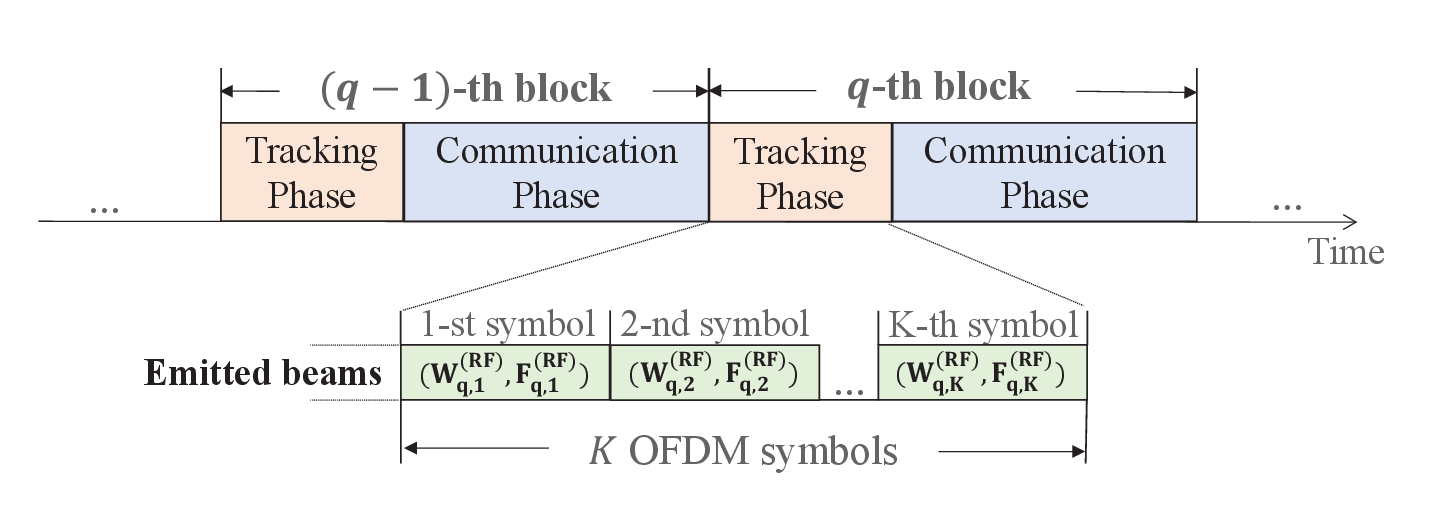}
\vspace{-0.3cm}
    \caption{The transmission protocol for the considered mmWave ISAC system.}
    \label{fig:protocol}
    \vspace{-0.5cm}
\end{figure}

Let $\bar{K}$ denote the number of OFDM symbols within a block, and $L$ denote the number of sub-carriers for each OFDM symbol. We divide each block into two phases - tracking phase that consists of $K<\bar{K}$ OFDM symbols and communication phase that consists of $\bar{K}-K$ OFDM symbols, as shown in Fig. \ref{fig:protocol}. In the tracking phase of each block, the BS transmits $B_{s}=KN_{\text{T}}^{\text{RF}}>I$ beams to track the $I$ targets over $K$ symbol durations. In the subsequent communication phase, the BS transmits $B_c=(\bar{K}-K)N_{\text{T}}^{\text{RF}}$ beams to deliver messages to the $U$ users over $\bar{K}-K$ symbol durations. Since mmWave communication has been widely studied, in this paper, we focus on the mmWave tracking task. 

During the tracking phase, the BS's transmit signal on the $l$-th sub-carrier of the $k$-th OFDM symbol in the $q$-th block can be expressed as
\vspace{-0.2cm}
\begin{equation}
\mathbf{x}_{q,k,l} = \sqrt{P} {\mathbf{W}^{\text{RF}}_{q,k}}\mathbf{W}^{\text{BB}}_{q,k,l} \mathbf{s}_{q,k,l}, \quad \forall q,k,l, \vspace{-0.2cm}
\end{equation}
where \(\mathbf{s}_{q,k,l}\in \mathbb{C}^{ N_{\text{T}}^\text{RF} \times 1}\) with $\mathbb{E}[\mathbf{s}_{q,k,l}{\mathbf{s}_{q,k,l}^H}]=\mathbf{I}_{N_{\text{T}}^\text{RF}}$ denotes the BS's transmit symbol vector on the $l$-th sub-carrier of the $k$-th OFDM symbol in the $q$-th block, $P$ denotes the transmit power, $\mathbf{W}^{\text{BB}}_{q,k,l} \in \mathbb{C}^{N_{\mathrm{T}}^{\mathrm{RF}} \times N_{\mathrm{T}}^{\mathrm{RF}}}$denotes the digital precoder on the $l$-th sub-carrier of the $k$-th OFDM symbol in the $q$-th block, and ${\mathbf{W}^{\text{RF}}_{q,k}}\in \mathbb{C}^{N_{\text{T}} \times N_{\text{T}}^\text{RF}}$ denotes the frequency-flat phase shifter-based analog precoder of the $k$-th symbol in the $q$-th block. Each element of \(\mathbf{W}^{\text{RF}}_{q,k}  \) satisfies the constant modulus constraint, i.e., \(
\left| \mathbf{W}^{\text{RF}}_{q,k}[n_i,n_j] \right|
=\frac{1}{\sqrt{N_{\text{T}}}}\) with $n_i=1,\ldots,N_\text{T}$ and $n_j=1,\ldots,N_{\text{T}}^{\text{RF}}$. Here, we consider that \(\mathbf{W}^{\text{RF}}_{q,k}  \) is designed by selecting $N_{\text{T}}^{\text{RF}}$ beams from a codebook
\vspace{-0.15cm}
\begin{equation}
\mathcal{W}_{\text{T}}=\left\{
\mathbf{w}_{\text{T}}^{(m)} \in\mathbb{C}^{N_\text{T} \times 1},
m = 1,\ldots,M_{\text{T}}
\right\}, \vspace{-0.15cm}
\end{equation}
where each beam \(\mathbf{w}_{\text{T}}^{(m)}\) is a pencil-like narrow beam towards some pre-designed azimuth angle $\bar{\theta}_m$ and elevation angle $\bar{\phi}_m$ \cite{chiang2021}. 

Moreover, let \( \mathbf{a}_{\text{T}}(\theta, \phi) \in \mathbb{C}^{N_\text{T} \times 1} \) and \( \mathbf{a}_{\text{R}}(\theta, \phi) \in \mathbb{C}^{N_\text{R} \times 1} \) denote the steering vectors of the transmit and receive UPAs of the BS towards the azimuth angle $\theta$ and elevation angle $\phi$, respectively, with the form $\mathbf{a}_{j}(\theta, \phi)= 
\mathbf{a}_{j}^{x}(\theta, \phi) \otimes \mathbf{a}_{j}^{y}(\theta, \phi)$ for $j\in \{\text{T},\text{R}\}$, where $\otimes $ denotes the Kronecker product, and

\vspace{-0.2cm}
\begin{small}
\begin{align}
\mathbf{a}_j^{x}(\theta, \phi) 
&= [
1, 
e^{j2\pi \frac{d_s}{\lambda} \sin \phi \cos \theta}, 
\cdots,
e^{j2\pi \frac{d_s}{\lambda} (N_j^{x}-1)\sin \phi \cos \theta}]^T,\nonumber \\
\mathbf{a}_{j}^{y}(\theta, \phi) 
&= [
1, e^{j2\pi \frac{d_s}{\lambda} \sin \phi \sin \theta}, \cdots,
e^{j2\pi \frac{d_s}{\lambda} (N_{j}^{y}-1)\sin \phi \sin \theta}]^T,\vspace{-0.2cm}
\end{align}
\end{small}with $d_s$ being the antenna spacing, and $\lambda$ being the carrier wavelength. In practical mmWave systems, the line-of-sight (LoS) channel model is widely adopted \cite{liu2022learningJSAC, Niu2015mmWaveSurvey}. Therefore, the round-trip channel matrix between the BS and target $i$ on the $l$-th sub-carrier during the $k$-th symbol in the $q$-th block, denoted by \(\mathbf{H}_{i,q,k,l}\in \mathbb{C}^{N_\text{R} \times N_\text{T}}\), can be modeled as
\vspace{-0.1cm}
\begin{align}\label{eq:Ht}
\mathbf{H}_{i,q,k,l}
&= \alpha_{i,q}e^{j2\pi \nu_{i,q} kT_0}e^{-j2\pi l\Delta f\tau_{i,q} } \nonumber \\
 &\times \mathbf{a}_{\text{R}}(\theta_{i,q}, \phi_{i,q}) \mathbf{a}_{\text{T}}^H(\theta_{i,q}, \phi_{i,q}),\vspace{-0.1cm}
\end{align}
where $\alpha_{i,q}=\sqrt{\frac{\lambda^2}{(4\pi)^3 d_{i,q}^4}}\zeta_{i,q}$ is the complex coefficient with $\zeta_{i,q}$ denoting the radar cross-section (RCS) of target $i$ in the $q$-th block, $\tau_{i,q}$ and $\nu_{i,q}$ denote the time delay and Doppler frequency between the BS and target $i$, respectively, $T_0$ denotes the OFDM symbol duration, and $\Delta f$ denotes the sub-carrier spacing. Then, during the $k$-th symbol in the $q$-th block, the echo signal received by the BS on the $l$-th sub-carrier is given as
\vspace{-0.1cm}
\begin{align} \label{signal}
\boldsymbol{\upsilon}_{q,k,l} 
= \sqrt{P}&{(\mathbf{F}^{\text{RF}}_{q,k}\mathbf{F}^{\text{BB}}_{q,k,l})}^H \sum_{i=1}^I \mathbf{H}_{i,q,k,l} {\mathbf{W}^{\text{RF}}_{q,k}}\mathbf{W}^{\text{BB}}_{q,k,l} \mathbf{s}_{q,k,l} \nonumber \\
&+{(\mathbf{F}^{\text{RF}}_{q,k}\mathbf{F}^{\text{BB}}_{q,k,l})}^H\mathbf{n}_{q,k,l}, \quad \forall q, k, l,
\end{align}
where \(\mathbf{n}_{q,k,l} \sim \mathcal{CN}(\mathbf{0}, \sigma^2\mathbf{I}_{N_{\text{R}}})\) denotes the additive white Gaussian noise (AWGN) vector at the BS with \(\sigma^2\) being the power, \(\mathbf{F}^{\text{BB}}_{q,k,l} \in \mathbb{C}^{N_{\text{R}}^\text{RF} \times N_{\text{R}}^\text{RF}}\) denotes the BS receive digital combiner on the $l$-th sub-carrier of the $k$-th symbol in the $q$-th block, and \(\mathbf{F}^{\text{RF}}_{q,k} \in \mathbb{C}^{N_{\text{R}}\times N_{\text{R}}^\text{RF}}\) denotes the BS receive phase shifter-based analog combiner. Similar to the analog precoder, $\mathbf{F}^{\text{RF}}_{q,k}$ is subject to \(
\left| \mathbf{F}^{\text{RF}}_{q,k}[n_i,n_j] \right|
=\frac{1}{\sqrt{N_{\text{R}}}}\) with $n_i=1,\ldots,N_{\text{R}}$ and $n_j=1,\ldots,N_{\text{R}}^{\text{RF}}$, and is designed by selecting $N_{\text{R}}^{\text{RF}}$ beams from the codebook
 \vspace{-0.15cm}
\begin{equation} \mathcal{W}_{\text{R}}=\left\{
\mathbf{w}^{(m)}_{\text{R}} \in\mathbb{C}^{N_{\text{R}}\times 1},
m = 1,\ldots,M_{\text{R}}
\right\}. \vspace{-0.15cm}
\end{equation}

By collecting the signals across all $L$ sub-carriers, the received signal at the BS of the $k$-th symbol in the $q$-th block is denoted as $\boldsymbol{\Upsilon}_{q,k} = [\boldsymbol{\upsilon}_{q,k,1}, \ldots, \boldsymbol{\upsilon}_{q,k,L}] \in \mathbb{C}^{N_\text{R}^{\mathrm{RF}} \times L}$. Furthermore, by stacking $\boldsymbol{\Upsilon}_{q,k}$ across $K$ OFDM symbols, the overall received signal matrix during the tracking phase in the $q$-th block is given as $\tilde{\boldsymbol{\Upsilon}}_{q} = \left[ \boldsymbol{\Upsilon}_{q,1}^T, \ldots, \boldsymbol{\Upsilon}_{q,K}^T \right]^T \in \mathbb{C}^{ K N_\text{R}^{\mathrm{RF}} \times L}$.

A conventional way to perform tracking based on $\tilde{\boldsymbol{\Upsilon}}_{q}$ is as follows. At the tracking phase of each block $q$, the BS first estimates the range and angular information of each target $i$ based on $\tilde{\boldsymbol{\Upsilon}}_{q}$, denoted by $\hat{\mathbf{z}}_{i,q}=[\hat{d}_{i,q}, \hat{\theta}_{i,q}, \hat{\phi}_{i,q}]^T$, where $\hat{d}_{i,q}, \hat{\theta}_{i,q}$ and $ \hat{\phi}_{i,q}$ are the estimated range, azimuth angle and elevation angle of target $i$, respectively. The parameter estimation problem can be well solved by classical methods such as maximum likelihood estimation (MLE) \cite{Stoica1990MLDOA} or data-driven approaches like deep neural networks \cite{Huang2018DOA}. Then, Kalman filter \cite{ekf} or data-driven based approaches \cite{kalmannet} can leverage the estimated distances and angles for tracking the targets. However, the above approaches are under the assumption that the echo signals from the targets are always strong enough for precise distance and angle estimation. In our considered mmWave systems, due to the analog beamformers, the transmit antenna array can merely emit pencil-like narrow beams, while the receive antenna array can merely receive signals from a narrow direction. Thus, the main challenge of the target tracking problem is how to maintain the narrow analog beams that keep aligned with the targets, known as beam tracking \cite{yi2024beam}, to obtain reliable location information. 

\vspace{-0.2cm}
\section{Problem Description and Transformer-based Solution }

\subsection{Problem Description}
This paper aims to predict the analog beams aligned with each target in each block to track the locations of these targets over time. Specifically, based on the trajectory of each target, the BS can predict $\{\mathbf{W}^{RF}_{q+1,k}, \mathbf{F}^{RF}_{q+1,k}\}_{k=1}^{K}$ for the next block $q+1$ by exploiting the historical sensing information up to block $q$. The analog beamformer predictor, denoted by $\mathcal{F}^{(q)}(\cdot)$, can be expressed as
\vspace{-0.1cm}
\begin{equation}\label{predictive beamforming}
\{\mathbf{W}^{\text{RF}}_{q+1,k}, \mathbf{F}^{\text{RF}}_{q+1,k}\}_{k=1}^{K}=\mathcal{F}^{(q)}(\mathcal{H}_{1:q}),\quad \forall q,\vspace{-0.1cm}
\end{equation}
where $\mathcal{H}_{1:q}=\{\{\hat{\mathbf{z}}_{i,\tau}\}_{i=1}^I,\{\mathbf{W}^{\text{RF}}_{\tau,k}, \mathbf{F}^{\text{RF}}_{\tau,k}\}_{k=1}^{K}\}_{\tau=1}^{q}$ denotes the historical sensing information that consists of the measurements of each target and designed analog beamformers in each historical block $\tau \in \{1,\ldots, q\}$. 

Let $\mathcal{B}^{(\text{tx})}_{q+1,k} \subset \mathcal{W}_{\text{T}}$ and $\mathcal{B}^{(\text{rx})}_{q+1,k} \subset \mathcal{W}_{\text{R}}$ denote the sets of predicted beams from the codebook $\mathcal{W}_\text{T}$ and $\mathcal{W}_\text{R}$ for designing $\mathbf{W}^{\text{RF}}_{q+1,k}$ and $ \mathbf{F}^{\text{RF}}_{q+1,k}$, respectively, with $|\mathcal{B}^{(\text{tx})}_{q+1,k}|=N_\text{T}^{\text{RF}}$ and $|\mathcal{B}^{(\text{rx})}_{q+1,k}|=N_\text{R}^{\text{RF}}$. The analog beamformer predictor in \eqref{predictive beamforming} can be transformed into a predictor for the beam subsets, denoted by $\tilde{\mathcal{F}}^{(q)}$, which can be expressed as
\vspace{-0.1cm}
\begin{equation}
\label{predictive_codeword}
\{\mathcal{B}^{(\text{tx})}_{q+1,k}, \mathcal{B}^{(\text{rx})}_{q+1,k}\}_{k=1}^K=\tilde{\mathcal{F}}^{(q)}(\tilde{\mathcal{H}}_{1:q}),\quad \forall q,\vspace{-0.1cm}
\end{equation}
where $\tilde{\mathcal{H}}_{1:q} = \{\{\hat{\mathbf{z}}_{i,\tau}\}_{i=1}^I,\{\mathcal{B}^{(\text{tx})}_{\tau,k}, \mathcal{B}^{(\text{rx})}_{\tau,k}\}_{k=1}^K\}_{\tau=1}^{q}$ represents the historical information of target measurements and selected beam subsets in each historical block $\tau \in \{1,\ldots, q\}$. 

Let $(\mathbf{w}^\ast_{\text{T}, i,q+1}, \mathbf{w}^\ast_{\text{R}, i,q+1})$ denote the optimal transmit and receive beam pair that is aligned with the target $i$ in the $q+1$-th block, which is defined as

\vspace{-0.2cm}
\begin{small}
\begin{align}
&(\mathbf{w}^\ast_{\text{T}, i,q+1}, \mathbf{w}^\ast_{\text{R}, i,q+1}) = \underset{\substack{\mathbf{w}_{\text{T}} \in \mathcal{W}_{\text{T}}, \\ \mathbf{w}_{\text{R}} \in \mathcal{W}_{\text{R}}}}{\arg\max} \nonumber \\
&\left| \mathbf{w}_{\text{R}}^{H} \mathbf{a}_{\text{R}}(\theta_{i,q+1}, \phi_{i,q+1}) \mathbf{a}_{\text{T}}^H(\theta_{i,q+1}, \phi_{i,q+1}) \mathbf{w}_{\text{T}} \right|^2 .
\end{align}
\end{small}

In the $q+1$-th block, we consider that the target $i$ will be tracked if the beamformers of the BS are aligned with the target during one OFDM symbol in the tracking phase. For a given symbol, beam alignment is equivalent to the condition that the optimal transmit and receive beam pair for the target is contained in the predicted beam subsets obtained in \eqref{predictive_codeword}. Let $\mathcal{A}_{i,q+1,k}$ denote the beam alignment event for target $i$ during the $k$-th symbol. Then, the tracked event of the target $i$ in the $q+1$-th block, denoted by  $\mathcal{E}_{i,q+1}^{\mathrm{(track)}}$, can be defined as 
\vspace{-0.2cm}
 \begin{equation}
      \label{track_event}
\mathcal{E}_{i,q+1}^{\mathrm{(track)}}
\triangleq
\bigcup_{k=1}^{K} \mathcal{A}_{i,q+1,k},\vspace{-0.1cm}
 \end{equation}
where
\vspace{-0.2cm}
\begin{small}
 \begin{equation}
\mathcal{A}_{i,q+1,k} \triangleq 
\left\{
\mathbf{w}^\ast_{\text{T}, i,q+1} \in \mathcal{B}^{(\text{tx})}_{q+1,k}
\right\}
\cap
\left\{
\mathbf{w}^\ast_{\text{R}, i,q+1} \in \mathcal{B}^{(\text{rx})}_{q+1,k}
\right\}.\vspace{-0.1cm}
\end{equation}
\end{small}

However, if beam alignment for target $i$ fails across all $K$ symbols, we consider that this target is lost in the $q+1$-th block, because the echo reflected from the target is too weak to extract accurate sensing information, i.e., $\hat{\mathbf{z}}_{i,q+1}$ is missing. Accordingly, the lost event for target $i$ in the $q+1$-th block, denoted by  $\mathcal{E}_{i,q+1}^{\mathrm{(lost)}}$, can be defined as
\vspace{-0.2cm}
 \begin{equation}
      \label{lost_event}
\mathcal{E}_{i,q+1}^{\mathrm{(lost)}}
\triangleq
\bigcap_{k=1}^{K} \mathcal{A}^c_{i,q+1,k},\vspace{-0.2cm}
 \end{equation}
where $\mathcal{A}^{c}_{i,q+1,k}$ denotes the complementary event of $\mathcal{A}_{i,q+1,k}$, corresponding to beam misalignment during the $k$-th symbol.
 
Therefore, our objective is to design $\tilde{\mathcal{F}}^{(q)}(\cdot)$ to maximize the probability that all $I$ targets can be successfully tracked in the next block $q+1$, such that we can achieve persistent tracking of all the targets. The optimization problem can be formulated as
\vspace{-0.1cm}
\begin{subequations}
\begin{align}
(\mathrm{P1}):\quad 
\max_{\tilde{\mathcal{F}}^{(q)}(\cdot)} 
\quad 
& \mathbb{P}\!\left(
\bigcap_{i=1}^{I} \mathcal{E}_{i,q+1}^{\mathrm{(track)}}
\,\middle|\,
\tilde{\mathcal{H}}_{1:q}
\right)
\label{eq:obj_P1} \\
\text{s.t.}\quad
&  \eqref{predictive_codeword},
\\
& |\mathcal{B}^{(\text{tx})}_{q+1,k}| = N_{\mathrm{T}}^{\mathrm{RF}}, 
|\mathcal{B}^{(\text{rx})}_{q+1,k}| = N_{\mathrm{R}}^{\mathrm{RF}}, \label{p1.1}
\\
& \mathcal{B}^{(\text{tx})}_{q+1,k} \subset \mathcal{W}_{\text{T}}, 
\mathcal{B}^{(\text{rx})}_{q+1,k} \subset \mathcal{W}_{\text{R}}. \label{p1.2}
 \vspace{-0.1cm}
\end{align}
\end{subequations}

It is observed that solving problem $(\mathrm{P1})$ via traditional optimization-based techniques is challenging due to the following two reasons. First, the mapping from historical sensing information to future beam subsets in \eqref{predictive_codeword} is highly non-linear, which limits the effectiveness of traditional KF-based tracking methods. Second, the tracking lost issue in the mmWave system implies that the historical sensing information $\tilde{\mathcal{H}}_{1:q}$ could be incomplete due to missing measurements, which challenges the robustness of the predictor in \eqref{predictive_codeword}, particularly in high-mobility scenarios with complex trajectories. Therefore, the design of $\tilde{\mathcal{F}}^{(q)}(\cdot)$ should have the capability to maintain accurate beam prediction even if the measurements from some targets are unavailable in the history, while adaptively adjusting the beam sweeping strategy in response to the tracking lost events.

\vspace{-0.2cm}
\subsection{Proposed Transformer-based Solution}
To tackle these challenges, in this paper, we propose to leverage the deep learning technique to solve the problem $(\mathrm{P1})$, where the powerful Transformer architecture is adopted. Specifically, the proposed Transformer-based tracking method operates in two modes depending on whether the tracking lost event occurs, which aims to realize accurate beam prediction and fast re-acquisition of lost targets simultaneously. The two modes are described as follows:
\subsubsection{Normal Tracking Mode (N-Mode)} In the $q$-th block, our system operates in the N-Mode if all targets are hit by the swept beams and there is no tracking lost event. In this mode, the objective is to predict the swept beam subsets for the $q+1$-th block based on historical trajectories up to the $q$-th block, such that all targets can be continuously tracked in the $q+1$-th block with a high probability. 
\subsubsection{Re-Acquisition Mode (R-Mode)} In the $q$-th block, the system operates in the R-Mode if there exist targets misaligned with all swept beams, resulting in the tracking lost events. The objective of this mode is to adjust future beam sweeping strategy by jointly exploiting the positive information from historical trajectories and the negative feedback from beam misalignment, such that the lost targets can be re-acquired as soon as possible, while the non-lost targets can be continuously tracked in the $q+1$-th block with a high probability. 
\begin{remark}
By separating the operation into the N-Mode and R-Mode, the proposed framework can adaptively learn the beam sweeping strategies for maintaining normal tracking and re-acquiring lost targets separately. Compared to the single-mode framework, this design can significantly reduce the difficulty of model training and improve the overall tracking robustness. 
\end{remark}

In the next section, we provide a detailed introduction of the proposed two-mode tracking framework.

\vspace{-0.15cm}
\section{Transformer-based Tracking Framework for the Single-Target Case}
To provide a comprehensive introduction on the proposed Transformer-based tracking framework, we start from the single-target case in this section, namely the design of $\text{m}^2$TrackFormer. The generalization of the proposed framework to the multi-target case, i.e., the $\text{m}^3$TrackFormer, will be discussed in Section V. Moreover, we consider a symmetric configuration where the number of antennas and RF chains for the transmitter and receiver are equal, i.e., $N_\text{T}=N_\text{R}=N$, $N_\text{T}^\text{RF}=N_\text{R}^\text{RF}=N^\text{RF}$. Due to the monostatic sensing, the optimal transmit beam and receive beam for the target are equal. Therefore, we only focus on predicting the optimal beam in the codebook $\mathcal{W}$ without distinguishing the transmit and receive sides\footnote{The extension to the asymmetric scenario is straightforward by predicting the optimal beams in the two different sets $\mathcal{W}_{\text{T}}$ and $\mathcal{W}_{\text{R}}$ simultaneously.}.

\begin{figure*}[!t]
  \centering
  \includegraphics[width=0.8\textwidth]{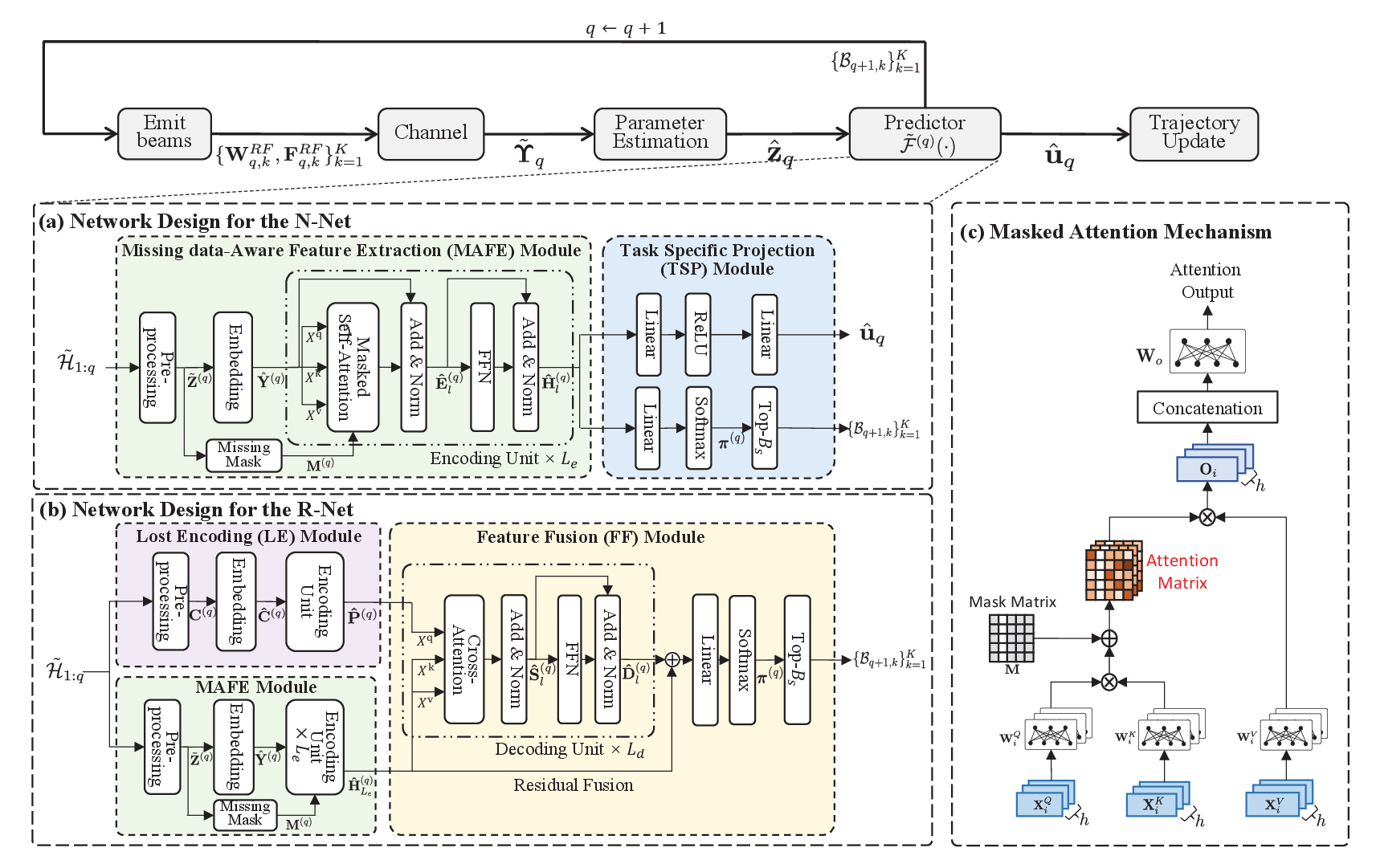} 
  \vspace{-0.4cm}
  \caption{The proposed Transformer-based tracking framework in the single-target case, namely $\text{m}^2$TrackFormer. In the predictor $\tilde{\mathcal{F}}^{(q)}(\cdot)$, the left upper branch illustrates the N-Net, where the historical sensing information is processed sequentially by the MAFE and TSP modules to predict future beam directions. The left lower branch shows the R-Net, in which the historical sensing information and feedback from the tracking lost events are processed via the MAFE and LE modules, respectively, followed by the feature fusion in the FF module to adjust future beam sweeping strategy. The right branch shows the multi-head Masked Attention Mechanism employed in the Self-Attention layers and Cross-Attention layers.}
  \label{fig:network}
  \vspace{-0.4cm}
\end{figure*}

In the proposed $\text{m}^2$TrackFormer, we design two networks to implement the tasks in  the N-Mode and R-Mode, respectively, as shown in Fig. \ref{fig:network}. Specifically, a Normal Tracking Network (N-Net), denoted by $\mathcal{F}^{(q)}_{\mathrm{N}}(\cdot; \Theta_{\mathrm{N}})$ with trainable parameters $\Theta_{\mathrm{N}}$, is proposed for the N-Mode to learn the beam sweeping strategy for persistent tracking, while a Re-Acquisition Network (R-Net), denoted by $\mathcal{F}_{\mathrm{R}}^{(q)}(\cdot; \Theta_{\mathrm{R}})$ with trainable parameters $\Theta_{\mathrm{R}}$, is proposed for the R-Mode to learn the beam sweeping strategy for target re-acquisition. In the following subsections, we detail the module design of each network, and introduce the policy for training the networks.

 \vspace{-0.3cm}
\subsection{Design of the N-Net}
For the N-Net, we aim to accurately predict the beam subset for the next block $q+1$ and localize the target for the current block $q$ based on the historical sensing information of the target up to block $q$. The N-Net is comprised by two modules: the Missing data-Aware Feature Extraction module (MAFE) and the Task-Specific Projection module (TSP). Specifically, the MAFE is designed to extract the target motion feature from the historical measurements, and the TSP is utilized to map the extracted high-dimensional feature to the prediction of beam subset and the estimation of target position. By leveraging the Self-Attention mechanism and a missing-aware attention mask, the N-Net achieves \textit{non-recursive} feature extraction directly from incomplete trajectories, thereby eliminating the need for data imputation and mitigating error propagation. These two modules contributing to the N-Net are detailed introduced as follows.

\subsubsection{MAFE module}
In the N-Mode, the sequence of historical measurements is directly used as the input to the MAFE module to extract temporal features of the target trajectory. To avoid overly long sequence, a sliding-window truncation strategy of fixed length $T_p$ is employed to retain only the measurements from the recent $T_p$ time blocks. Consequently, for block $q$, the network input is constructed as $\hat{\mathbf{Z}}^{(q)}=[\hat{\mathbf{z}}_{q-T_p+1},\ldots, \hat{\mathbf{z}}_{q}]^T \in \mathbb{R}^{T_p\times D_{in}}$, where $D_{in}=3$ is the dimension of measurements consisting of the estimated range, azimuth and elevation angle. If tracking lost occurs at some historical time blocks, zero-padding is applied for these positions to ensure dimension consistency. Then, we append a learnable \textit{Prediction Token} at the end of the temporal dimension of $\hat{\mathbf{Z}}^{(q)}$ to enable the prediction for the next block $q+1$, which extends the temporal dimension to $T_e = T_p + 1$. The extended sequence is denoted as $\tilde{\mathbf{Z}}^{(q)} = [\hat{\mathbf{Z}}^{(q)}; \mathbf{0}] \in \mathbb{R}^{T_e \times D_{in}}$. Subsequently, $\tilde{\mathbf{Z}}^{(q)}$ is mapped into a unified feature space with dimension $D$ via a learnable embedding layer, which can be formulated as $  \hat{\mathbf{Y}}^{(q)}  = \phi_{\text{proj}}(\tilde{\mathbf{Z}}^{(q)}) + \mathbf{P}^{(pos)}\in \mathbb{R}^{T_e \times D }$, where $\phi_{\text{proj}}(\cdot): \mathbb{R}^{T_e \times D_{in}} \!\rightarrow\! \mathbb{R}^{T_e \times D }$ is a linear projection layer, and $\mathbf{P}^{(pos)}$ represents the sinusoidal positional encoding to preserve temporal order \cite{Vaswani2017Attention}.

To mitigate the impact of missing measurements in feature extraction, a \textit{Missing-Aware Attention Mask} is employed, denoted by $\mathbf{M}^{(q)}  \in \{\eta, 0\}^{T_e \times T_e}$. Here, $\eta$ represents a large negative scalar (e.g., $\eta = -10^9$) that drives the attention weights of masked positions to zeros after the Softmax operation in the Attention mechanism. Specifically, the entries of $\mathbf{M}^{(q)}$ are defined as $
\mathbf{M}^{(q)}[:,j]=\eta\cdot \mathbf{1}_{T_e}$ if the $j$-th row of $\tilde{\mathbf{Z}}^{(q)}$ is zero-padded, and $
\mathbf{M}^{(q)}[:,j]=\mathbf{0}$ otherwise, $\forall j \in\{1,\ldots, T_e\}$. The function of the missing mask is to suppress information propagation from blocks corresponding to missing measurements and the appended token, thereby ensuring that feature extraction is performed exclusively on valid measurements and preventing error propagation. 

Subsequently, the embedded sequence $\hat{\mathbf{Y}}^{(q)} $ and the missing mask $\mathbf{M}^{(q)} $ served as inputs to $L_e$ stacked Encoding Units for feature extraction. In the $l$-th Encoding Unit $(l = 1, \ldots, L_e)$, the input $\hat{\mathbf{H}}^{(q)} _{l-1}$ (with \( \hat{\mathbf{H}}_0^{(q)}  = \hat{\mathbf{Y}}^{(q)}  \)) is first processed by a Self-Attention (SA) layer, which captures globally temporal correlations in the historical trajectory by computing attention scores across multiple attention heads. Specifically, the Multi-Head Attention mechanism, denoted by $\text{Attn}(\cdot)$, can be expressed as \cite{Vaswani2017Attention}  
\vspace{-0.15cm}
\begin{equation}
\label{multi-head attention}
\text{Attn}(\mathbf{X}^K, \mathbf{X}^Q, \mathbf{X}^V, \mathbf{M})  =  \text{Concat}(\mathbf{O}_1, \ldots, \mathbf{O}_h)\mathbf{W}_o,\vspace{-0.2cm}
\end{equation}
\vspace{-0.2cm}
\begin{align}
\mathbf{O}_i=  \text{Softmax}& \left( \frac{( \mathbf{X}^Q\mathbf{W}_i^Q )( \mathbf{X}^K\mathbf{W}^K_i)^T}{\sqrt{D_i}} + \mathbf{M} \right)\left(\mathbf{X}^V\mathbf{W}_i^V\right) , \nonumber \\
& \qquad \qquad \qquad \qquad i = 1, \ldots, h, \vspace{-0.3cm}
\end{align}
where $\mathbf{X}^Q$, $\mathbf{X}^K$ and $\mathbf{X}^V$ denote the query, key and value matrix, respectively, $\text{Concat}(\cdot)$ denotes the matrix splicing operation along the feature dimension, $\mathbf{O}_i$ is the output of the Attention operation for the $i$-th Attention head, $\mathbf{W}_o \in \mathbb{R}^{D \times D}$, $\mathbf{W}_i^Q ,\mathbf{W}_i^K,\mathbf{W}_i^V\in \mathbb{R}^{D \times D_i}$ are learnable projection matrices, $D_i=D/h$ is the dimension of $i$-th Attention head.

The output of the SA layer is processed through a Feed-Forward Network (FFN). To facilitate gradient flow and stabilize training, residual connections and Layer Normalization (LN) are applied after both the SA layers and FFNs. Specifically, the feature extraction in the $l$-th Encoding Unit can be expressed as
\vspace{-0.15cm}
\begin{equation} 
\label{encoder1}
\hat{\mathbf{E}}_l^{(q)}  = \text{LN}\left( \text{Attn}(\hat{\mathbf{H}}_{l-1}^{(q)} ,\hat{\mathbf{H}}_{l-1}^{(q)} , \hat{\mathbf{H}}_{l-1}^{(q)} , \mathbf{M}^{(q)} ) + \hat{\mathbf{H}}_{l-1} ^{(q)} \right) ,\vspace{-0.15cm}
\end{equation}
\vspace{-0.15cm}
\begin{equation}
\mathbf{\hat{H}}_l^{(q)}  = \text{LN}(\text{FFN}(\mathbf{\hat{E}}_l^{(q)} ) + \mathbf{\hat{E}}_l^{(q)} ) . \label{encoder2}\vspace{-0.15cm}
\end{equation}
\subsubsection{TSP module} Upon obtaining the final latent representation $\hat{\mathbf{H}}_{L_e}^{(q)} \in \mathbb{R}^{T_e \times D}$ from the Encoding Units, the TSP module is employed to decouple the feature processing for the primary beam prediction task and an auxiliary localization refinement task using the Multi-Task Learning (MTL) strategy \cite{mtl}. For the beam prediction task, the prediction token at temporal index $T_e$ is passed through a classifier $\phi_{\text{cls}}(\cdot)$ to yield the predicted probability vector $\boldsymbol{\pi}^{(q)} = [\pi_{q}^{(1)}, \ldots,\pi_{q}^{(M)}]^T \in [0,1]^{M}$, which can be expressed as $ \boldsymbol{\pi}^{(q)} = \phi_{\text{cls}}(\hat{\mathbf{H}}_{L_e}^{(q)} [T_e,:])$, where $\pi_{q}^{(m)}$ represents the predicted likelihood that the $m$-th beam in the codebook will align with the target in the next block $q+1$. Finally, a Top-$B_s$ selection strategy is employed to construct the sensing beam subset $\{\mathcal{B}_{q+1,k}\}_{k=1}^K$ by selecting the $B_s$ indices with the highest probabilities in $\boldsymbol{\pi}^{(q)}$. 

Moreover, the location refinement task is performed to refine the target’s estimated location for the current block $q$ (index by $T_p$) via a MLP layer, which can be expressed as $ \hat{\mathbf{u}}_{q} = \phi_{\text{reg}}(\hat{\mathbf{H}}_{L_e}^{(q)} [T_p,:])$, where $\phi_{\text{reg}}(\cdot)$ denotes the refinement function. This MTL-based auxiliary task offers two advantages: first, it leverages historical trajectory to filter measurement noise, thereby enhancing localization precision; second, it helps learn a precise target mobility representation, which can also improve the beam prediction task.

\vspace{-0.3cm}
\subsection{Design of the R-Net}
In the tracking lost scenario, the beam prediction problem becomes more difficult because the absence of range and angular measurements of the target in block $q$ inevitably increases the uncertainty of the future target trajectory. Based on empirical results, we find that it is hard to train a well-performed model by only exploiting the historical measurement sequence $\hat{\mathbf{Z}}^{(q)}$ by masking out the missing entries. Motivated by the fact that the misaligned beam directions in the lost events also carry implicit mobility information of the target about where the target is unlikely to be located, we propose to extract the target mobility information in the R-Net from two features: the \textit{Motion Feature} obtained from the historical measurement sequence, and the \textit{Lost Feature} obtained from the tracking lost events. Specifically, the R-Net contains three modules: 1) the MAFE module for motion feature extraction; 2) the Lost Encoding (LE) module for tracking lost feature extraction; 3) the Feature Fusion (FF) module for combining the two types of features to adjust the beam sweeping prediction via a \textit{residual fusion} strategy. In the following, we detail the three modules contributed to the R-Net.

\begin{remark}
It is worth noting that adjusting the beam sweeping strategy in the proposed framework does not guarantee immediate target re-acquisition in the next block. As a result, a target may remain to be lost over multiple consecutive time blocks. Nevertheless, the proposed scheme is capable of gradually adjusting the re-acquisition strategy by exploiting the lost events observed in each lost block. To ensure robustness and avoid indefinite re-acquisition attempts, a tracking failure will be declared if the number of consecutive tracking lost events exceeds a predefined maximum threshold $T_{\text{max}}$.
\end{remark}

\subsubsection{MAFE module} The MAFE module processes the historical measurement sequence $\hat{\mathbf{Z}}^{(q)} $ with the same architecture and parameters as in the N-Net, and outputs the latent motion feature $\hat{\mathbf{H}}_{L_e}^{(q)}\in \mathbb{R}^{T_e \times D} $ of the target. 

\subsubsection{LE module} The objective of the LE module is to exploit the implicit mobility information from the misaligned beam directions to enable target re-acquisition. Since the target may be lost for multiple consecutive time blocks, let $q_l$ denote the index of the last block where the target was successfully tracked. In the LE module, we consider all the tracking lost events from time block $q_l+1$ to $q$ to adjust the beam sweeping strategy for the $q+1$-th block. Specifically, let $\mathcal{B}_{\tau}=\cup_{k=1}^K \mathcal{B}_{\tau,k}
$ denote the set of all the swept beams over the $K$ symbols in the $\tau$-th block, where $\tau=q_l+1,\ldots, q$. Therefore, the sequence of lost events, which serves as the input to the LE module, is denoted as $\mathcal{S}_B^{(q)}=\{\mathcal{B}_{q_l+1},\ldots, \mathcal{B}_{q}\}$. Note that the size of $\mathcal{S}_B^{(q)}$ is dynamic, which is determined by the block duration of consecutive tracking lost, i.e., $T^{(q)}_{\text{lost}}=q-q_l+1$. 

Since the raw beam set $\mathcal{B}_{\tau}$ cannot be directly processed by the neural network, we encode each set into a \textit{multi-hot lost vector}, denoted by $\mathbf{c}_\tau  \in \{0,1\}^{M}$. The $m$-th entry of $\mathbf{c}_\tau$ is defined as $c_{\tau,m} =1$ if $\mathbf{w}^{(m)} \in \mathcal{B}_\tau$, and $c_{\tau,m} =0$ otherwise. By stacking these vectors along the temporal dimension, we obtain the feedback matrix \( \mathbf{C}^{(q)} =[\mathbf{c}_{q_l+1}, \ldots, \mathbf{c}_{q}]^T\in \{0,1\}^{ T^{(q)}_{\text{lost}} \times M} \). Similar to the MAFE module, a learnable Prediction Token is appended at the end of \( \mathbf{C}^{(q)}\), yielding \( \tilde{\mathbf{C}}^{(q)}=[\mathbf{C}^{(q)}; \mathbf{0}]\in \{0,1\}^{T^{(q)}_{l} \times M}\) with an extended temporal dimension of $T^{(q)}_{l}=T^{(q)}_{\text{lost}}+1$. Subsequently, $\tilde{\mathbf{C}}^{(q)} $ is projected to the latent feature space by an embedding layer  $\phi_{\text{emb}}(\cdot): \{0,1\}^{T^{(q)}_{l} \times M  } \!\rightarrow\! \mathbb{R}^{T^{(q)}_{l} \times D }$, followed by the positional encoding to preserve the temporal information. The resulting high-dimensional latent feature can be expressed as $
\hat{\mathbf{C}}^{(q)}  = \phi_{\text{emb}}(\tilde{\mathbf{C}}^{(q)} )  + \text{P}^{(pos)} \in \mathbb{R}^{T^{(q)}_{l} \times D}$. 

To capture the temporal correlations within the lost events across time blocks, $\hat{\mathbf{C}}^{(q)}$ is further processed through an Encoding Unit following the operations in \eqref{encoder1}-\eqref{encoder2}. In this context, the query, key, and value matrices are all set to $\hat{\mathbf{C}}^{(q)}$. Furthermore, the mask matrix $\bar{\mathbf{M}}^{(q)}$ is an all-zero matrix because all the entires are valid. Consequently, the output of the Encoding Unit, denoted by $\mathbf{\hat{P}}^{(q)}\in \mathbb{R}^{T^{(q)}_{l} \times D}$, encapsulates the learned \textit{lost feature} from the tracking lost events.


\subsubsection{FF module} The outputs from the MAFE and LE modules, i.e., $\hat{\mathbf{H}}_{L_e}^{(q)} $ and $\mathbf{\hat{P}}^{(q)}$, are then fused in the FF module to adjust the beam prediction result based on the tracking lost events. Motivated by the fusion capability of the Cross-Attention (CA) mechanism, we employ $L_d$ Decoding Units, each consisting of a CA layer and a FFN with LN and residual connection, to extract the cross-sequence information from the two features. Formally, the process of the $l$-th Decoding Unit (\( l = 1, \ldots, L_d \)), can be expressed as
\vspace{-0.15cm}
\begin{equation}
\label{decoder1}
\hat{\mathbf{S}}_l^{(q)}  = \text{LN}(\text{Attn}(\hat{\mathbf{D}}_{l-1}^{(q)} , \hat{\mathbf{H}}_{L_e}^{(q)} , \hat{\mathbf{H}}_{L_e}^{(q)} , \hat{\mathbf{M}}^{(q)} ) + \hat{\mathbf{D}}_{l-1}^{(q)} ),\vspace{-0.15cm}
\end{equation}
\vspace{-0.15cm}
\begin{equation}
\label{decoder2}
    \hat{\mathbf{D}}_l^{(q)}  = \text{LN}(\text{FFN}(\hat{\mathbf{S}}_l^{(q)} ) + \hat{\mathbf{S}}_l^{(q)} ),\vspace{-0.15cm}
\end{equation}
where \( \hat{\mathbf{D}}_{0}^{(q)} =\mathbf{\hat{P}}^{(q)}  \), $\hat{\mathbf{M}}^{(q)}\in \{\eta, 0\}^{T_{l}^{(q)} \times T_e}$ denote the Cross-Attention mask. Similar to the Missing Mask in the MAFE, the entries of $\hat{\mathbf{M}}^{(q)}$ are defined as $
\hat{\mathbf{M}}^{(q)}[:,j]=\eta\cdot \mathbf{1}_{T_l^{(q)}}$, if the $j$-th row of $\tilde{\mathbf{Z}}^{(q)}$ is zero-padded, and $
\hat{\mathbf{M}}^{(q)}[:,j]=\mathbf{0}$ otherwise, $\forall j \in\{1,\ldots, T_e\}$.

After \(L_d \) Decoding Units, we extract the latent feature from the appended token in $\hat{\mathbf{D}}_{L_d}^{(q)}$, denoted as \(\hat{\mathbf{q}}_{\text{feed}}^{(q)} 
= \hat{\mathbf{D}}_{L_d}^{(q)} [T^{(q)}_{l},:]\in \mathbb{R}^{D}\), as the latent representation for the next block $q+1$ learned from the tracking lost events. Then, we propose an innovative \textit{residual fusion} strategy that incorporates the lost feature as a \textit{corrective refinement} rather than a replacement of the original motion representation for beam prediction. Specifically, let \(\hat{\mathbf{q}}_{\text{mo}}^{(q)}=\hat{\mathbf{H}}_{L_e}^{(q)} [T_{e},:] \in \mathbb{R}^{D}\) denote the original motion feature for block $q+1$ extracted from the MAFE module. We formulate the refined motion feature $\hat{\mathbf{q}}_{\text{re}}^{(q)} \in \mathbb{R}^{D}$ as
\vspace{-0.1cm}
\begin{equation}
\label{fusion}
\hat{\mathbf{q}}_{\text{re}}^{(q)} 
= \hat{\mathbf{q}}_{\text{mo}}^{(q)} 
+ \hat{\mathbf{q}}_{\text{feed}}^{(q)} .\vspace{-0.1cm}
\end{equation}
Note that this residual fusion strategy requires the LE and FF modules to learn only the deviation induced by the lost events on the motion feature. Subsequently, $\hat{\mathbf{q}}^{(q)} _{\text{re}}$ is fed into the classifier $\phi_{\text{cls}}(\cdot)$ to yield the final beam probability vector $\boldsymbol{\pi}^{(q)}$, followed by the Top-$B_s$ selection strategy on $
    \boldsymbol{\pi}^{(q)}$ to obtain $\{\mathcal{B}_{q+1,k}\}_{k=1}^K$. 
\vspace{-0.2cm}
\subsection{Training Policy}
The training policy also has an essential effect on the performance of the proposed $\text{m}^2$TrackFormer. However, the key challenge is that the distribution of tracking lost events, which serves as the input for the R-Net, is dependent on the real-time tracking failures of the N-Mode and cannot be naturally captured by a static offline dataset. To address this, we propose a two-phase supervised training strategy, inspired by the Dataset Aggregation strategy \cite{Ross2011DAgger}, to actively collect the target lost samples for robust re-acquisition training. 

\subsubsection*{Dataset Construction and Pre-processing}
First, we construct a raw trajectory dataset $\mathcal{D}_{\mathrm{raw}}=\{\hat{\mathbf{Z}}^{(n)}, \mathbf{U}^{(n)}, \mathbf{m}^{*(n)}\}_{n=1}^{N_d}$, containing $N_d$ trajectories. Each trajectory $n$ consists of $Q$ samples, comprising noisy measurements $\hat{\mathbf{Z}}^{(n)}=\left[\hat{ \mathbf{z}}_1^{(n)}, \ldots, \hat{\mathbf{z}}_{Q}^{(n)} \right]$, true target coordinates $\mathbf{U}^{(n)}=\left[ \mathbf{u}_1^{(n)}, \ldots, \mathbf{u}_{Q}^{(n)} \right]$, and optimal beam indices $\mathbf{m}^{*(n)}=\left[ m_1^{*(n)}, \ldots, m_{Q}^{*(n)} \right]$. This dataset is then partitioned into two subsets: a basic subset $\mathcal{D}_{\mathrm{base}}$ used for pre-training the N-Net, and a held-out subset $\mathcal{D}_{\mathrm{sim}}$ used for generating tracking lost samples and training the R-Net.

\subsubsection*{Two-Phase Training Strategy}
Now we introduce the two-phase training strategy to facilitate the training of the N-Net and R-Net. In Phase 1, the MAFE and TSP modules are pre-trained on $\mathcal{D}_{\mathrm{base}}$ to learn the mapping from historical observations to the future optimal beams and current location. To enhance the model's robustness on capturing long-term temporal dependencies, we adopt a multi-token prediction training strategy \cite{multitoken}, where the modules are trained to simultaneously predict optimal beams for multiple blocks in the future. The total loss is a weighted combination of the Cross-Entropy (CE) loss and the Mean Squared Error (MSE) loss, which is given by
\vspace{-0.2cm}
\begin{equation}
\label{loss1}
\mathcal{L}_{1} = \mathcal{L}_{\mathrm{CE}} + \lambda \mathcal{L}_{\mathrm{MSE}},
\vspace{-0.2cm}
\end{equation}
where $\lambda$ is a weighting coefficient. The individual loss components are defined as
\vspace{-0.2cm}
\begin{align}
\mathcal{L}_{\mathrm{CE}} &= - \frac{1}{N_s} \sum_{n=1}^{N_s} \sum_{\tau=q+1}^{q+\delta_ q}\sum_{m=1}^{M} y_{n,\tau}^{(m)} \log \pi_{n,\tau}^{(m)}, \\ 
\mathcal{L}_{\mathrm{MSE}} &= \frac{1}{N_s} \sum_{n=1}^{N_s} \| \hat{\mathbf{u}}_{n,q} - \mathbf{u}_{n,q} \|_2^2,
\vspace{-0.3cm}
\end{align}
where $N_s$ denotes the batch size, $M$ is the codebook size, $\delta_q$ denotes the prediction horizon, and $y_{n,\tau}^{(m)}$ is the binary ground-truth indicator which equals 1 if the $m$-th beam is the optimal beam for the $n$-th sample in block $\tau$, and 0 otherwise.

In Phase 2, we aim to train the LE and FF modules in the R-Net. First, we generate the tracking lost samples via executing the well-trained MAFE and TSP modules on $\mathcal{D}_{\mathrm{sim}}$. During this process, we monitor the tracking status in each time block. If a tracking lost event occurs, we record the swept beam indices in the lost event as a sample, denoted by $\tilde{\mathcal{B}}_{lost}$, together with the corresponding trajectory information in $\mathcal{D}_{sim}$. The collected data forms an augmented re-acquisition dataset $\tilde{\mathcal{D}}_{\mathrm{reacq}}= \{ \mathcal{D}_{sim}, \tilde{\mathcal{B}}_{lost} \}$. Subsequently, the parameters in the LE and FF modules are trained based on $\tilde{\mathcal{D}}_{\mathrm{reacq}}$, while the parameters of the MAFE and TSP modules are frozen. The loss function in this phase is the CE loss, given by
\vspace{-0.2cm}
\begin{equation}
\label{loss2}
\mathcal{L}_{2} = - \frac{1}{|\tilde{\mathcal{D}}|} \sum_{n=1}^{|\tilde{\mathcal{D}}|} \sum_{\tau=q+1}^{q+\delta_ q}\sum_{m=1}^{M} y_{n,\tau}^{(m)} \log \pi_{n,\tau}^{(m)}.
\vspace{-0.1cm}
\end{equation}





\vspace{-0.2cm}
\section{Transformer-based Tracking Framework for the Multi-Target Case}

\begin{figure*}[t]
    \centering
\includegraphics[width=0.8\linewidth]{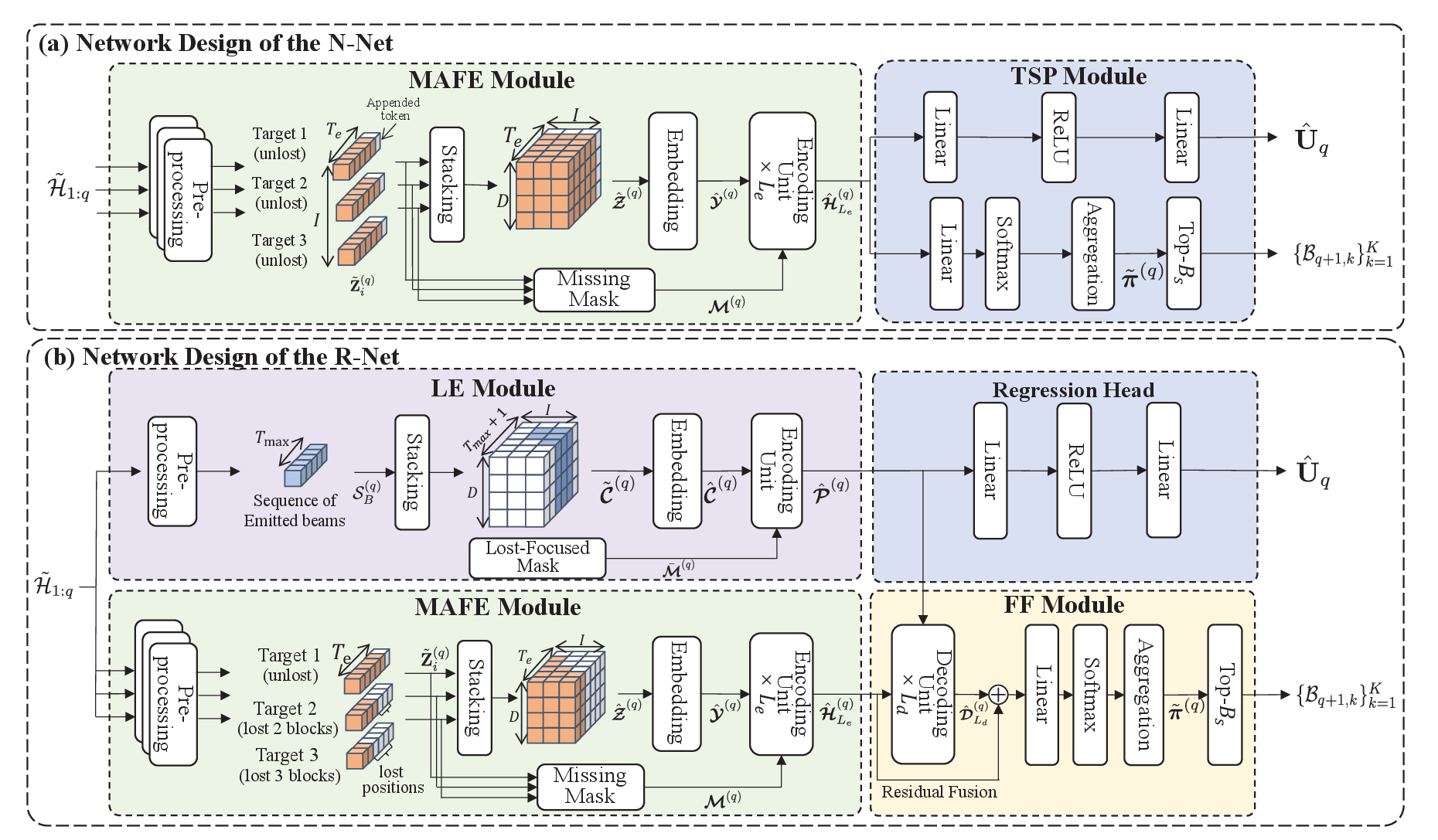}
\vspace{-0.3cm}
    \caption{The diagram for the proposed $\text{m}^3$Trackformer.}
    \label{multi overview}
    \vspace{-0.5cm}
\end{figure*}

In this section, we extend to investigate the general tracking framework for the multi-target tracking scenario, where there are \( I > 1 \) moving targets to be tracked over the tracking duration of \( Q \) blocks. Compared with the single-target scenario, the beam prediction task in the multi-target case becomes more challenging due to the following two reasons. First, the model is required to predict beam subsets that can cover the possible directions of all targets, so that the emitted beams could be aligned with all targets rather than concentrating on being aligned with one target. Second, the numbers of lost and unlost targets vary dynamically over time due to the occurrence of tracking lost and re-acquisition events. Therefore, the designed framework should accommodate dynamic variations on the number of targets. 

To address these challenges, we propose an $\text{m}^3$TrackFormer, which follows the similar idea of the $\text{m}^2$TrackFormer. For the first challenge, we introduce a feature aggregation layer, which combines individual target features to formulate a joint beam sweeping strategy under a shared beam budget, which adaptively balances the number of swept beams to re-acquire lost targets while maintaining accurate tracking for unlost targets. For the second challenge, we propose a scalable architecture that leverages the parameter-sharing strategy and the inherent parallel processing capability in Transformer to process an arbitrary number of targets simultaneously without the need for model retraining, thereby achieving low inference latency in real-time tracking scenarios. The network design of the $\text{m}^3$TrackFormer is illustrated in Fig. \ref{multi overview}. In the following, we detail the key differences between the $\text{m}^3$TrackFormer and $\text{m}^2$TrackFormer.
\vspace{-0.3cm}
\subsection{Design of the N-Net}
The N-Net of $\text{m}^3$TrackFormer also applies the MAFE and TSP modules for the beam prediction and the target localization tasks. To accommodate multiple targets, the historical sequences of each target $i \in \mathcal{I}$, denoted by $\hat{\mathbf{Z}}_i^{(q)}$, are stacked along a newly introduced \textit{target dimension} to form a unified input tensor $\hat{\boldsymbol{\mathcal{Z}}}^{(q)}\in \mathbb{R}^{I \times T_e \times D_{in}   }$. Moreover, we extend the missing-aware attention mask in $\text{m}^2$TrackFormer to a three-dimensional mask tensor, denoted by $\boldsymbol{\mathcal{M}}^{(q)} \in \{\eta, 0\}^{I\times T_{e} \times T_{e}}$, to mitigate the impact of missing measurements in feature extraction for each target. Specifically, we set ${\boldsymbol{\mathcal{M}}}^{(q)}[i, :,j]=\eta\cdot \mathbf{1}_{T_e}$ if the $j$-th row of $\hat{\mathbf{Z}}_i^{(q)}$ is zero-padded, and ${\boldsymbol{\mathcal{M}}}^{(q)}[i,:,j]=\mathbf{0}$ otherwise, $\forall i\in\{1,\ldots, I\}$, $j \in \{1,\ldots, T_e\}$.

To address the multi-target tracking problem in $(\mathrm{P1})$, the $\text{m}^3$TrackFormer is designed to satisfy the properties of permutation equivariance and invariance. Specifically, any permutation of the target indices should not affect the overall beam sweeping strategy for the next time block (invariance), while the extracted target-wise motion features and localization results should be permuted accordingly (equivariance). To achieve the permutation equivariance, we adopt the parameter-sharing strategy \cite{Ravanbakhsh2017Equivariance} on the MAFE and TSP modules, where the same network layers are shared across the target dimension. Under this design, the motion feature of each target is extracted using the same function weights, which ensures that the extracted features transform equivariantly with respect to permutations of the target indices. As a result, the proposed architecture is scalable to the number of targets and enables direct deployment by reusing the weights of $\text{m}^2$TrackFormer trained on the single-target dataset, thereby avoiding costly model retraining when the number of targets varies. Therefore, taking $\hat{\boldsymbol{\mathcal{Z}}}^{(q)}$ and $\boldsymbol{\mathcal{M}}^{(q)}$ as inputs, the MAFE module yields the learned motion feature tensor, denoted by $\tilde{\boldsymbol{\mathcal{H}}}_{L_e}^{(q)} \in \mathbb{R}^{I \times T_e \times D}$. Accordingly, the MLP layer in the TSP module maps the features to the refined target positions, represented by the matrix $\hat{\mathbf{U}}^{(q)}\in \mathbb{R}^{I\times D_{out}}$, where $D_{out}=3$ denotes the output dimension, and the $i$-th row represents its refined coordinate location for target $i$, i.e., $\hat{\mathbf{u}}_{i,q}$. 

To further achieve the permutation invariance property for the beam prediction task, a feature aggregation layer is introduced in the TSP module to combine the features of all the targets. Specifically, let $\tilde{\boldsymbol{\pi}}^{(q)} \in \mathbb{R}^{ M}$ denote a beam score vector that yields the joint beam prediction result for tracking all the targets, which is given by 
\vspace{-0.2cm}
\begin{equation}
\tilde{\boldsymbol{\pi}}^{(q)} 
= \phi_{\text{agg}} \left( \phi_{\text{cls}}\!\left(\tilde{\boldsymbol{\mathcal{H}}}_{L_e}^{(q)}[:,T_e,:]\right)\right),\vspace{-0.2cm}
\end{equation}
where $\phi_{\text{agg}}(\cdot)$ denotes the feature aggregation function implemented via the non-parametric sum pooling function over the target dimension. A larger value of $\tilde{\boldsymbol{\pi}}^{(q)}[m]$ indicates that the $m$-th beam in the codebook $\mathcal{W}$ has a high confidence to be aligned with one target in the next time block. Finally, the BS can select the $B_s$ beams with the highest scores in $\tilde{\boldsymbol{\pi}}^{(q)}$ to form the candidate beam subset $\{\mathcal{B}_{q+1,k}\}_{k=1}^K$ for the next block $q+1$. 

\vspace{-0.3cm}
\subsection{Design of the R-Net}
The R-Net of $\text{m}^3$TrackFormer preserves the MAFE, LE, and FF modules, while further incorporating the regression head from the TSP module to refine the localization of unlost targets. The key challenge in the R-Net lies in the time-varying number of lost targets per block and their heterogeneous duration of tracking lost, which complicates the design of the LE and FF modules for multi-target re-acquisition. Since the MAFE module maintains the same architecture and parameters as in the N-Net for motion feature extraction of all the targets by \eqref{encoder1}-\eqref{encoder2} regardless of their tracking status, in the following we focus on the specialized design of the LE and FF modules. 

The LE module employs a \textit{dynamic padding} strategy along both the temporal and target dimensions to facilitate parallel processing of all the lost targets in a single forward pass. Let $\mathcal{I}_{\mathrm{unlost}}^{(q)}$ and $\mathcal{I}_{\mathrm{lost}}^{(q)}$ denote the sets of unlost and lost targets in the $q$-th block, respectively, and recall that $T_{\text{max}}$ is the threshold of maximum consecutive tracking-lost duration. In the R-Net, we construct a unified tracking-lost tensor $\tilde{\boldsymbol{\mathcal{C}}}^{(q)} \in \mathbb{R}^{I \times (T_{\text{max}}+1) \times M}$ as the input to the LE module, in which the entries corresponding to the unlost targets and the non-lost blocks for the lost targets are padded with zeros. Specifically, let $\mathbf{c}_{q_j}$ denote the multi-hot lost vector that encodes the swept beam indices in block $q_j$, where $q_j=q-T_{\text{max}}+j$ for $j\in\{1,\ldots, T_{\text{max}}+1\}$. Then, the entries of the $\tilde{\boldsymbol{\mathcal{C}}}^{(q)}$ are defined as $\tilde{\boldsymbol{\mathcal{C}}}^{(q)}[i, j,:]=\mathbf{c}_{q_j}$ if target $i \in \mathcal{I}_{\mathrm{lost}}^{(q)}$ is lost in the $\tau_j$-th block, and $\tilde{\boldsymbol{\mathcal{C}}}^{(q)}[i, j,:]=\mathbf{0}$ otherwise. Moreover, we introduce a \textit{lost-focused mask}, denoted by $\bar{\boldsymbol{\mathcal{M}}}^{(q)} \in \{\eta, 0\}^{I\times (T_{\text{max}}+1) \times (T_{\text{max}}+1)}$, which forces the attention mechanism in the LE module to concentrate on the lost events of each lost target while prevents information leakage to the unlost targets and non-lost blocks. Specifically, we set $\bar{\boldsymbol{\mathcal{M}}}^{(q)}[i, :,j]=\mathbf{0}$ if the target $i \in \mathcal{I}_{\mathrm{lost}}^{(q)}$ is lost in the $\tau_j$-th block, and $\bar{\boldsymbol{\mathcal{M}}}^{(q)}[i, :,j]=\eta \cdot \mathbf{1}$ otherwise, $\forall i\in\{1,\ldots, I\}$, $j\in\{1,\ldots, T_{\text{max}}+1\}$. Based on the constructed tensor inputs $\tilde{\boldsymbol{\mathcal{C}}}^{(q)}$ and $\bar{\boldsymbol{\mathcal{M}}}^{(q)}$, the subsequent embedding layer and the Encoding Unit extract the lost feature $\tilde{\boldsymbol{\mathcal{P}}}^{(q)} \in \mathbb{R}^{I \times (T_{\text{max}+1}) \times D}$ from the historical tracking lost events of each lost target. 

Subsequently, the FF module processed the motion feature $\tilde{\boldsymbol{\mathcal{H}}}_{L_e}^{(q)}$ and the lost feature $\tilde{\boldsymbol{\mathcal{P}}}^{(q)}$ of all the targets by \eqref{decoder1}-\eqref{fusion}, yielding the beam score vector $\tilde{\boldsymbol{\pi}}^{(q)} 
= \phi_{\text{agg}} \left( \phi_{\text{cls}}\!\left(\tilde{\boldsymbol{\mathcal{H}}}_{L_e}^{(q)}[:,T_e,:]+\tilde{\boldsymbol{\mathcal{D}}}_{L_e}^{(q)}[:,T_{\text{max}+1},:]\right)\right)$. Since both the motion features and lost features preserve the same target ordering under permutation, the subsequent decoding and fusion operations remain permutation equivariant. In particular, for unlost targets, the zero-padding strategy in the lost feature ensures that their motion features remain unaffected in the Decoding Units and residual fusion operation.


\vspace{-0.1cm}
\section{Numerical Experiments}

\subsection{Experiment Setup}

In this section, we provide numerical experiments to verify the effectiveness of the proposed tracking framework. The BS is located at \(\mathbf{p}_{BS}=[0, 0, 10]^T\) in meters (m) and is equipped with uniform transmit and receive UPAs, each with \(N_T=N_R=32\times32\) antennas and parallel to the $(x, y)$-plane. The discrete Fourier transform (DFT) codebook \cite{dft} with size $M=1024$ is adopted, and each transmit or receive beamforming vector has to be selected from this codebook. 

Similar to \cite{uavmoving, Han2025ActiveSensing}, the mobility of each target $i$ is modeled as \(\mathbf{u}_{i,q+1} = \mathbf{u}_{i,q} + \mathbf{v}_{i,q} \Delta T\), where \( \mathbf{u}_{i,q} = [ x_{i,q}, y_{i,q}, z_{i,q}]^T\) and \( \mathbf{v}_{i,q} = [v_{i,q}^x, v_{i,q}^y, v_{i,q}^z]^T\) denote the coordinate location and velocity vector of the target $i$ in the $q$-th block, respectively, and \( \Delta T=0.1 \)~s represents the time duration of a block. Here, we assume $ v_{i,q}^x = v_{i} \cos(\beta_{i,q}), v_{i,q}^y=v_{i} \sin(\beta_{i,q})$ and $v_{i,q}^z=0$, where the target speed $v_{i}$ is assumed to remain constant within a trajectory but varies across different targets and ranges from $[10,35]$~m/s. The moving direction in the horizontal plane $\beta_{i,q}$ follows a dynamic model over time, given by $\beta_{i,q+1} = \beta_{i,q} + \Delta \beta$, where $\Delta \beta \sim \mathcal{U}(-20^{\circ}, 20^{\circ})$. The initial moving direction is selected as $\beta_1 \sim \mathcal{U}(0, 2\pi)$. We assume that the initial altitude of the targets is randomly selected between 50~m and 80~m, i.e., $z_1 \sim \mathcal{U}(50,80)$~m, where the AWGN is introduced on the z-coordinate of the target in each block, i.e., $z_{i,q+1}=z_{i,q}+\Delta z$, with $\Delta z\sim \mathcal{N}(0,\sigma_z^2)$. A dataset of $N_d=20,000$ trajectories is generated, each with a tracking duration of $T=50$~s, and is split into training, validation, and evaluation subsets with proportions of 80\%, 10\%, and 10\%, respectively. Moreover, the network is implemented using PyTorch with the \textit{AdamW} optimizer for training, with a learning rate of $3\times10^{-4}$ and $50$ training epochs. For the hyperparameters, we set $L_e = 3$, $L_d = 2$, $D = 256$, $h = 8$, and the batch size to 256\footnote{Our code can be found in \nolinkurl{https://github.com/ltk722/Transformer-based-mmWave-tracking}.}.

 \vspace{-0.3cm}
\subsection{Baseline Models}
To demonstrate the effectiveness of the proposed Transformer-based framework, the following baseline methods are considered:
\begin{itemize}
    \item \textbf{KF-based Scheme}: This approach employs the Extended Kalman Filter (EKF) \cite{heath} to predict targets' positions over time and select the Top-$B_s$ beams whose steering directions minimize the mean squared error (MSE) with respect to the predicted locations. 
    \item \textbf{RNN-based Scheme} \cite{Burghal2019MLBeamTracking}: This approach utilizes a standard RNN to perform beam tracking. In the event of a tracking lost, the RNN adopts a data imputation strategy by freezing the state update and propagating the last valid hidden state to predict beams for subsequent blocks.
    \item \textbf{Two-mode LSTM-based Scheme}: This approach constructs an advanced sequence-to-sequence architecture consisting of two cascaded LSTM sub-networks \cite{jiang2022lidarWCL}. The primary LSTM encodes the historical trajectory for beam prediction and localization, while the secondary LSTM is activated in cases of tracking lost by taking the latent features from the primary LSTM and the lost events as input to predict beams for re-acquisition.

    \item \textbf{Single-mode Transformer-based Scheme}: This approach implements a decoder-only Transformer architecture \cite{Vaswani2017Attention}. It is a single mode framework and utilizes one Transformer network for accomplishing both normal target tracking and lost target re-acquisition. 
\end{itemize}

Under all schemes, a tracking failure is declared if the targets remain lost after $T_{\max}=5$ consecutive time blocks.
\vspace{-0.2cm}
\subsection{Performance Evaluation in the single-target scenario}
In this subsection, we evaluate the tracking performance of the proposed tracking scheme in the single-target scenario. We adopt the successful tracking probability and the average tracking duration as the performance metrics. Specifically, the successful tracking probability is defined as \(P_{\text{S}}=N_{\text{tracked}}/N_{\text{total}}\), where \(N_{\text{total}}\) is the total number of test trajectories and \(N_{\text{tracked}}\) is the number of trajectories under which the target is still tracked after $T$~s. The average tracking duration is defined as $T_{avg}=(1/N_\text{total}) \sum_{n=1}^{N_\text{total}} T_n$, where $T_n$ denotes the tracking time for which the $n$-th target remains tracked within the total period of $T$~s.

\subsubsection*{1) Performance over Tracking Time} 
\begin{figure}[t]
    \centering
\includegraphics[width=0.75\linewidth]{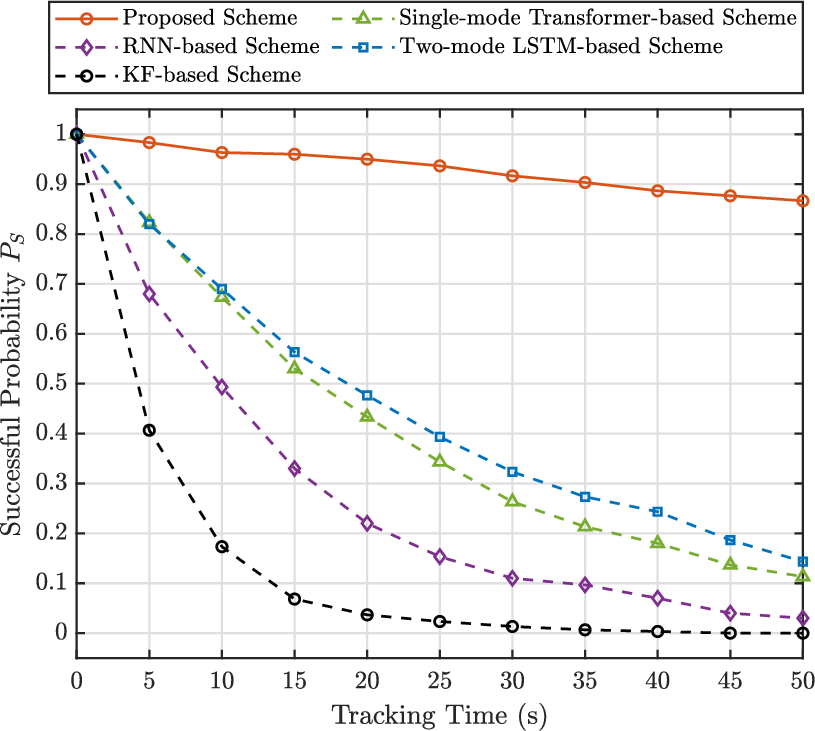}
 \vspace{-0.2cm}
    \caption{Tracking Performance of successful probability versus tracking time.}
    \label{result1}
    \vspace{-0.4cm}
\end{figure}

In Fig. \ref{result1}, we evaluate the successful tracking probability $P_S$ of various schemes versus the tracking time. The number of beams for tracking in each block is set to $B_s=4$. It is observed that our proposed scheme consistently achieves the highest successful probability as the tracking time increases. Specifically, the KF-based scheme suffers from severe performance degradation due to its inefficiency to model highly non-linear trajectories. Under the learning-based schemes, the RNN fails to provide reliable tracking because the frequent tracking lost events leads to error accumulation and eventually results in tracking failure. Although the Two-mode LSTM-based scheme attempts to enable re-acquisition, its improvement is still limited due to error accumulation. In contrast, the proposed scheme is able to maintain the tracking process in a much larger tracking interval. The performance gain is attributed to two main factors. First, the masked attention-empowered N-Mode achieves superior capability of feature extraction to avoid prediction error accumulation. Second, the re-acquisition mechanism in the R-Mode effectively re-acquire the lost target by exploiting the lost events, thereby increasing the tracking duration compared to the Single-mode Transformer-based scheme.

\subsubsection*{2) Performance over Number of Tracking Beams} 
\begin{figure}[t]
  \centering
  \subfigure[Performance of successful probability]
  {\includegraphics[width=.35\textwidth]{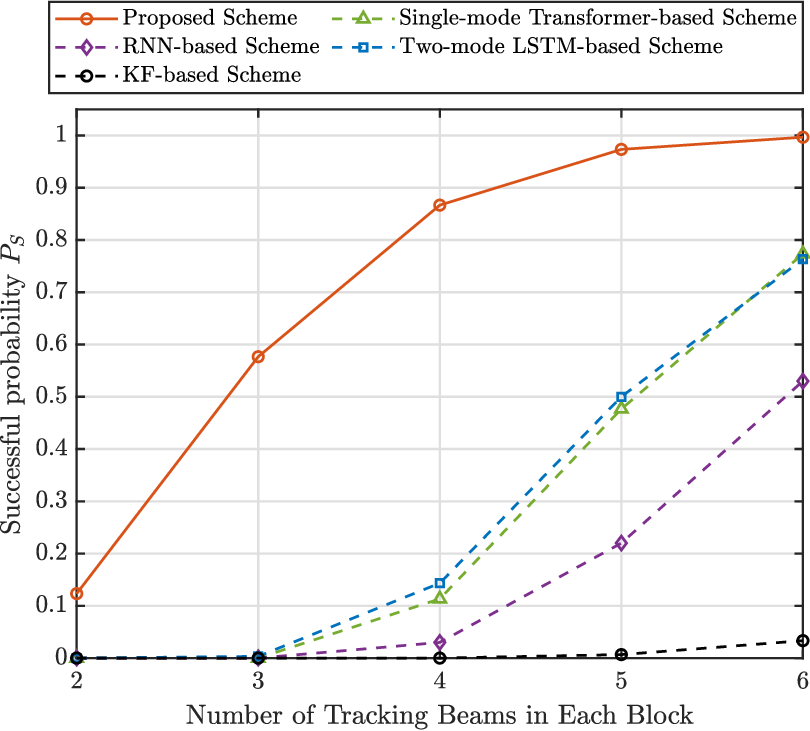}}\hspace{1cm}
  \subfigure[Performance of average tracking duration]
  {\includegraphics[width=.35\textwidth]{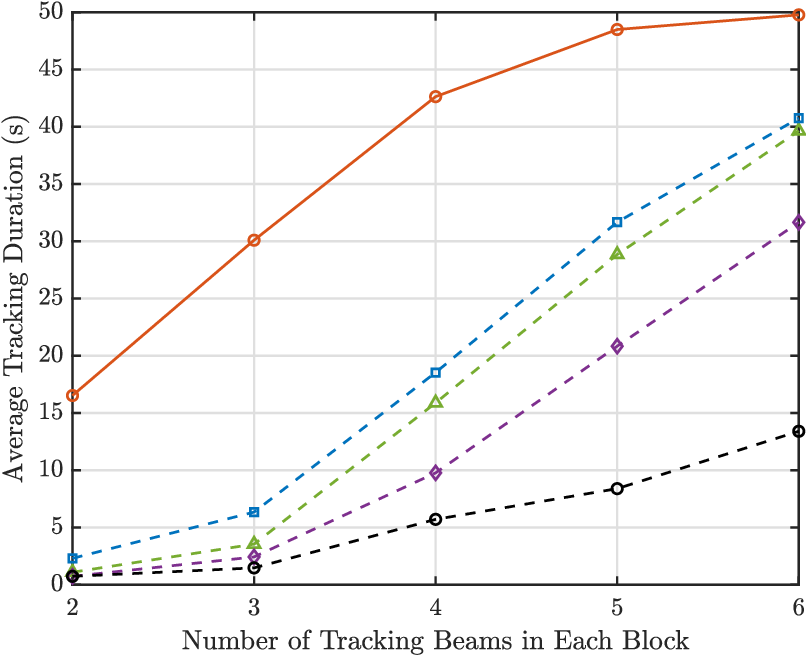}}
  \vspace{-0.3cm}
  \caption{Tracking performance versus the number of tracking beams.}\label{result2}
\end{figure}

In Fig. \ref{result2}, we evaluate the impact of the number of tracking beams per block, i.e., $B_s$, on the successful probability and the average tracking duration, respectively. It is observed that the proposed scheme achieves the highest successful probability and the longest tracking duration across all values of $B_s$. Specifically, when $B_s$ is small, all schemes exhibit limited performance on maintaining long-term tracking over 50~s due to insufficient spatial coverage of the limited tracking beams. However, the proposed scheme demonstrates superior robustness, achieving an average tracking duration over 15~s, while the benchmark schemes fail to maintain tracking for more than 5~s. As $B_S$ increases, the proposed scheme exhibits a rapid performance improvement. Notably, with as few as four beams, it achieves a successful probability exceeding 90\% and an average tracking duration of over 40~s. This result demonstrates the efficiency of the proposed framework on maintaining robust long-term tracking with minimal beam sweeping overhead. 

%
%

\subsubsection*{3) Performance over Target Speed} 
\begin{figure}[t]
    \centering
\includegraphics[width=0.7\linewidth]{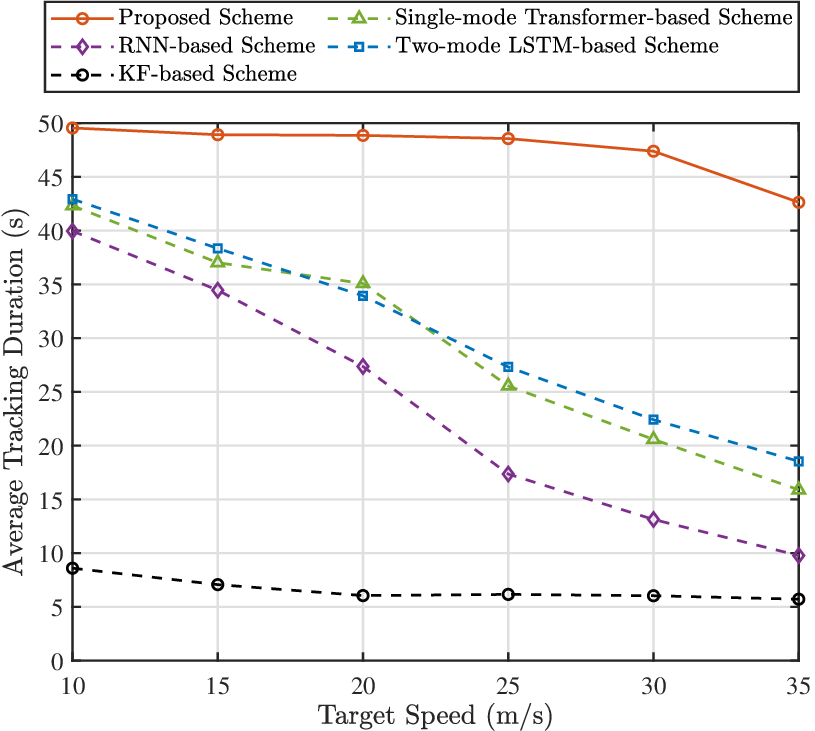}
 \vspace{-0.3cm}
    \caption{Average tracking duration versus target speed.}
    \label{result3}
    \vspace{-0.4cm}
\end{figure}

In Fig. \ref{result3}, we evaluate the impact of target speed on the average tracking duration with $B_s=4$ and $T=50~s$. It is observed that the tracking performance of all schemes degrades as the target speed increases. This degradation is because of the fact that higher mobility introduces greater stochasticity and non-linearity into the target's trajectory, thereby complicating the precise prediction of beam alignment. However, the proposed scheme demonstrates robustness to target mobility. Specifically, the average tracking duration decreases by less than 10~s when the target speed increases from the low-mobility case ($v=10$~m/s) to the high-mobility case ($v=35$~m/s). In contrast, other learning-based schemes suffer a more than 50\% degradation. Notably, compared to the Single-mode Transformer-based scheme, the proposed solution achieves a substantial performance gain ranging from 15\% (low mobility) to over 130\% (high mobility), which validates the critical role of the proposed R-Mode in facilitating effective target re-acquisition under high mobility scenarios.
 
\subsubsection*{4) Performance of Trajectory Estimation} 
\begin{figure}[t]
    \centering
\includegraphics[width=0.75\linewidth]{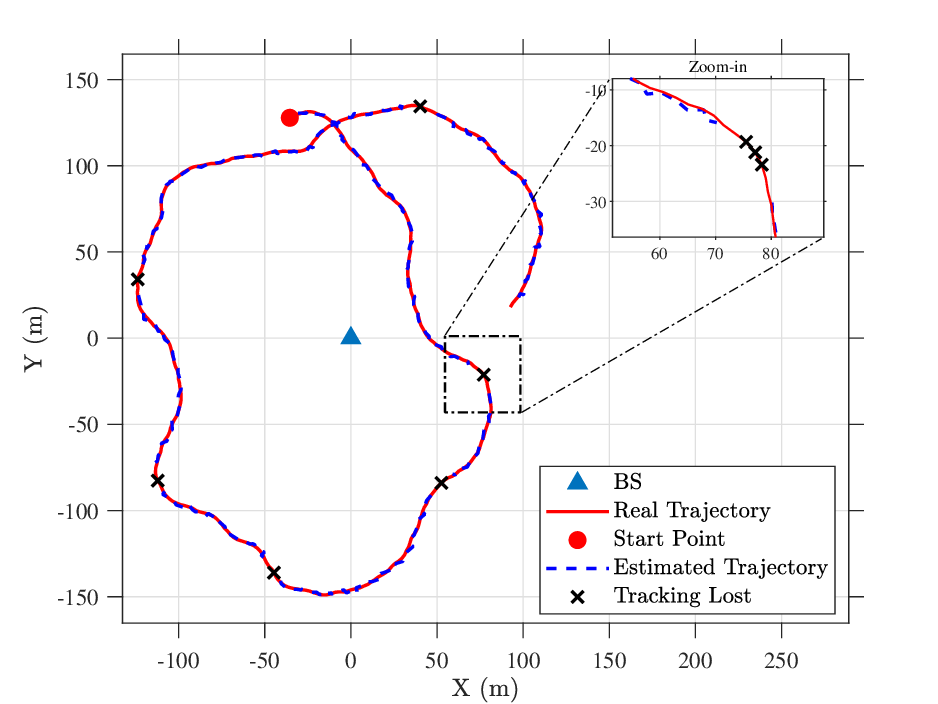}
 \vspace{-0.3cm}
    \caption{Illustration of the real and estimated trajectories.}
    \label{result4}
    \vspace{-0.5cm}
\end{figure}

In Fig. \ref{result4}, we illustrate the actual trajectory and the corresponding estimated locations of the target from the proposed scheme. It is observed that the tracking lost issue occurs particularly in regions where the target experiences rapid motion and the optimal beam direction changes abruptly. Although the target trajectory is complex, the proposed scheme is able to re-acquire the target location within a few blocks, and effectively track the trajectory over the entire tracking duration. 

\vspace{-0.3cm}
\subsection{Performance Evaluation in the Multi-Target Scenario}
In this subsection, we evaluate the tracking performance of the proposed framework in the multi-target scenario. As a benchmark, we consider a fully decoupled strategy that decomposes the multi-target tracking problem into $I$ independent single-target sub-tasks. 
Compared to the joint beam sweeping strategy in $\text{m}^3$TrackFormer, the benchmark processes each target sequentially with an individual sweeping strategy. The total beams for tracking in each time block are evenly divided among all targets, with $\lfloor B_s / I \rfloor$ beams to track each target. 

\begin{figure}[t]
  \centering
  \subfigure[Impact of the number of beams]
  {\includegraphics[width=.23\textwidth]{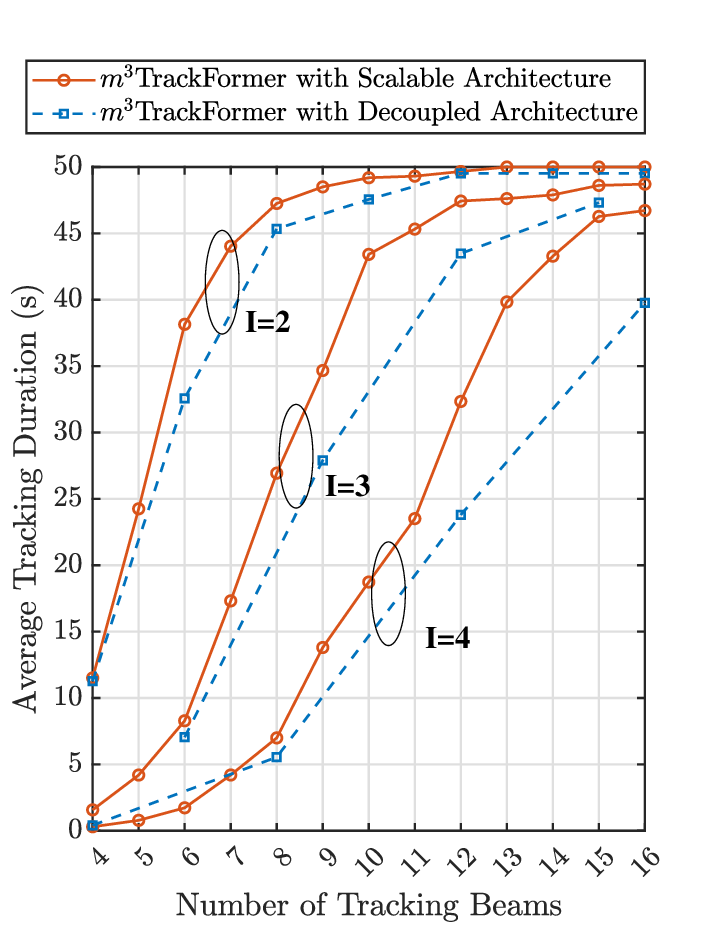}}\hspace{0.1cm}
  \subfigure[Impact of the number of targets]
  {\includegraphics[width=.23\textwidth]{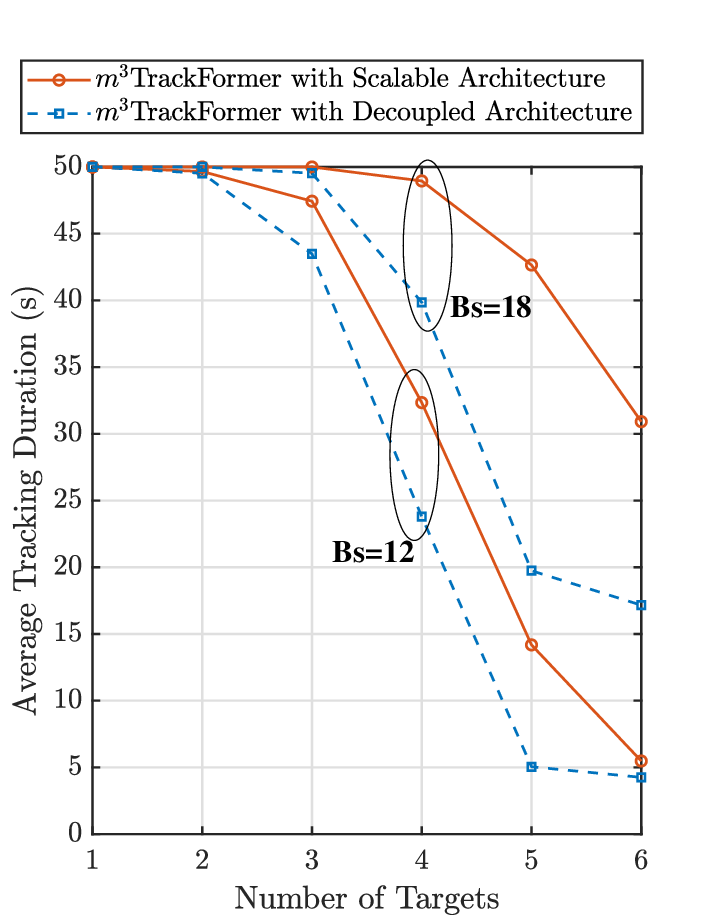}}
  \vspace{-0.3cm}
  \caption{Multi-target tracking performance of average tracking duration.}\label{result6_7}
  \vspace{-0.6cm}
\end{figure}

In Fig. \ref{result6_7}(a), we evaluate the performance of average tracking duration versus the number of tracking beams and the number of targets under the proposed scheme and the benchmark. It is observed that the performance of the proposed scheme achieves longer tracking duration than the benchmark in various number of beams, which demonstrates that the proposed joint beam sweeping strategy is effective on adaptively allocating the limited number of tracking beams to re-acquire the lost targets while maintaining normal tracking for unlost targets. In Fig. \ref{result6_7}(b), we further investigate the impact of the number of targets. It can be seen that the performance degrades as the number of targets increases, due to increased competition for limited beam resources. Nevertheless, the proposed scheme exhibits higher robustness in such scenarios, maintaining a substantially longer tracking duration than the benchmark. 
%

\begin{figure}[t]
    \centering
\includegraphics[width=0.8\linewidth]{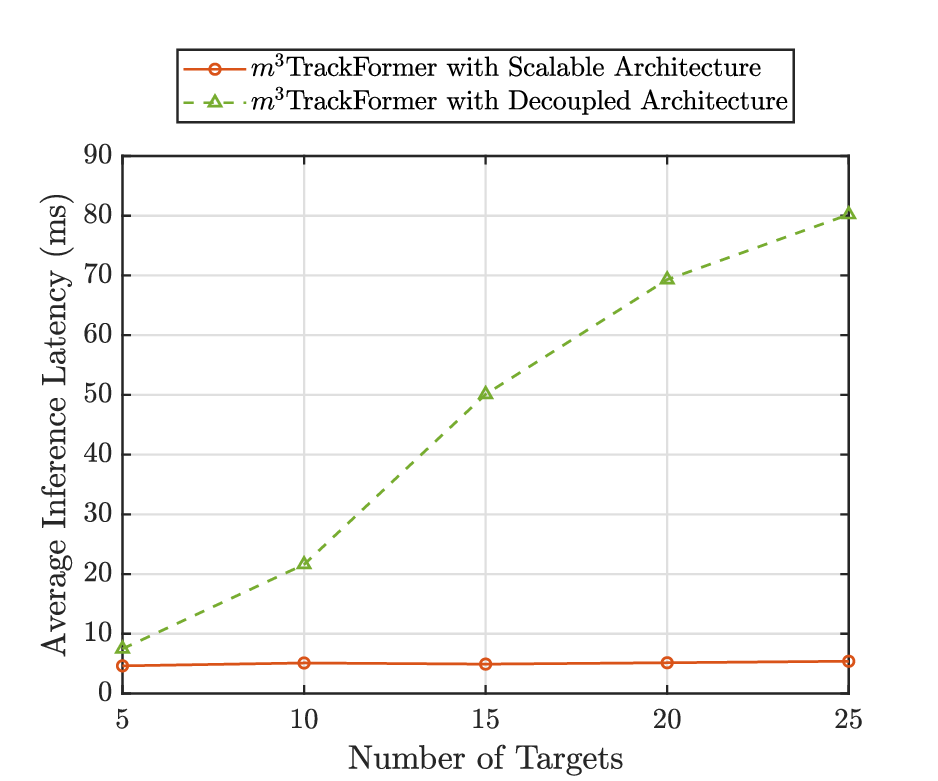}
 \vspace{-0.2cm}
    \caption{Average inference latency versus the number of targets.}
    \label{result8}
    \vspace{-0.6cm}
\end{figure}
In Fig. \ref{result8}, we evaluate the average inference latency per block over the number of targets. The simulations are conducted on an NVIDIA GeForce RTX 4060-Ti GPU. It is observed that the inference latency of $\text{m}^3$TrackFormer consistently remains below 10~milliseconds (ms) even as the number of targets increases. This efficiency comes from the inherent parallel processing capability on the Transformer backbone to process the features of all targets simultaneously within a single forward pass, which fully leverages the parallelism of the GPU. In contrast, the benchmark scheme exhibits a linear growth in latency, leading to significant computational overhead in dense scenarios. This result demonstrates that the proposed $\text{m}^3$TrackFormer satisfies the low-latency requirements of 6G ISAC applications and validates its feasibility for practical real-time deployment. 

\vspace{-0.2cm}
\section{Conclusion}
In this paper, we proposed a robust two-mode Transformer-based multi-target tracking framework for mmWave ISAC systems. When all the targets are hit by the swept beams, the framework operated in the N-Mode for realizing target tracking and beam prediction by the N-Net, in which the masked self-attention mechanism of Transformer is employed to extract global motion features of each target directly from incomplete historical trajectories. When the tracking lost event occurs due to beam misalignment, the framework switched to the R-Mode, in which the R-Net fused the motion features and negative feedback from beam misalignment to adjust the future beam sweeping strategy for target re-acquisition. Numerical results demonstrated that the proposed framework significantly outperforms benchmark schemes in terms of successful tracking probability and average tracking duration with low inference latency, which demonstrated its effectiveness and robustness for real-time multi-target tracking in the mmWave ISAC system.


\bibliographystyle{IEEEtran}

\bibliography{IEEEabrv,ref}

\end{document}